\documentclass{aa}  

\usepackage{comment} 
\usepackage{graphicx}
\usepackage{caption,subcaption}

\usepackage{amsmath}
\usepackage{txfonts}
\usepackage{hyperref}
\hypersetup{
    colorlinks,
    linkcolor={blue},
    citecolor={blue},
    urlcolor={blue}
}

\usepackage[noabbrev,capitalize,nameinlink]{cleveref}
\defcitealias{2025arXiv250905243P}{Paper~I}

\begin{document} 

   \title{Rethinking mass transfer: A unified semianalytical framework for circular and eccentric binaries}
   
   \subtitle{II. Orbital evolution due to nonconservative mass transfer}

   \author{A. Parkosidis\inst{1} \and 
           S. Toonen \inst{1} \and 
           E. Laplace \inst{2,3,1} \and 
           F. Dosopoulou \inst{4}
          }

   \institute{Anton Pannekoek Institute for Astronomy, University of Amsterdam, Amsterdam 1098 XH, The Netherlands\\
              \email{a.parkosidis@uva.nl} 
              \and
              Institute of Astronomy, KU Leuven, Celestijnenlaan 200D, B-3001 Leuven, Belgium
              \and 
              Leuven Gravity Institute, KU Leuven, Celestijnenlaan 200D, box 2415, 3001 Leuven, Belgium
              \and
             School of Physics and Astronomy, Cardiff University, Cardiff, CF24 3AA, United Kingdom
              }

   \date{Received November 11, 2025; accepted December 21, 2025}
 
  \abstract{Although mass transfer (MT) has been studied primarily in circular binaries, observations show that it also occurs in eccentric systems. We investigate orbital evolution during nonconservative MT in eccentric orbits, a process especially relevant for binaries containing compact objects (COs). We
  examined four angular momentum loss (AML) modes: Jeans, isotropic reemission, orbital-AML, and $L_2$ mass loss, with the last mode being the most efficient AML mode. For a fixed AML mode and accretion efficiency, orbital evolution is correlated: orbits either widen while becoming more eccentric, or shrink while circularizing. Jeans mode generally yields orbital widening and eccentricity pumping, whereas $L_2$ mass loss typically leads to orbital shrinkage and eccentricity damping. Isotropic reemission and orbital-AML show an intermediate behavior. Adopting isotropic reemission, we demonstrate that eccentric MT produces compact binaries that merge via gravitational waves (GW) within a Hubble time, whereas the same systems would instead merge during MT under traditional modeling. We further show that, in eccentric orbits, the gravitational potential at $L_2$ becomes lower than at $L_1$ across a wide range of mass ratios and eccentricities, naturally linking eccentricity to $L_2$ mass loss. Eccentric MT may therefore lead to the formation of the circumbinary disks observed around eccentric post-red-giant-branch and post-asymptotic-giant-branch systems. Since interacting binaries containing COs are frequently eccentric, $L_2$ mass loss offers a new robust pathway to orbital tightening during eccentric MT, contributing to the formation rate of GW sources. This model can treat orbital evolution due to conservative and nonconservative MT in arbitrary eccentricities, with applications ranging from MT on the main sequence to GW progenitors.}
    
    \keywords{binaries: close -- binaries: general -- celestial mechanics -- stars: kinematics and dynamics -- stars: mass-loss}

   \maketitle
%

\section{Introduction}\label{sec:one}

The orbital evolution of binary systems is shaped by various physical processes. Among these mechanisms, Roche-lobe overflow (RLOF) has long been understood as a key process that determines the evolution and final fate of mass-transferring binaries. Numerical models traditionally neglect orbital eccentricity during RLOF, assuming circular orbits or that tidal forces universally circularize orbits before the onset of RLOF \citep[e.g.,][]{1996A&A...309..179P,2002MNRAS.329..897H,2003ASPC..303..290P,2008ApJS..174..223B,2012A&A...546A..70T}. However, observations of interacting binaries with nonzero eccentricities \citep{1999AJ....117..587P,2005A&AT...24..151R}, theoretical studies predicting weak tides \citep[e.g.,][]{2009MNRAS.400L..20E,2022ApJ...933...25P}, and applications of more recent tidal formalisms \citep{2021MNRAS.503.5569V,2025ApJ...984..137D} challenge this assumption. Moreover, in recent years, a new picture has emerged in which eccentricity is a common feature of post-interaction wide binaries  \citep{2009Natur.462.1032M,2011Natur.478..356G,2016A&A...586A.158J,2018AJ....155..144K,2019A&A...626A.127J,2020A&A...641A.163V,2022A&A...658A.122M,2024MNRAS.52711719Y,2024MNRAS.529.3729S, 2024ApJ...962...70K,2025A&A...701A...9M} suggesting that RLOF may not only preserve but in some cases even enhance eccentricities \citep[][]{2007ApJ...667.1170S,2009ApJ...702.1387S,2019ApJ...872..119H,2025arXiv250905243P}.

Despite growing interest in the role of eccentric mass transfer (MT) in both analytical \citep{2019ApJ...872..119H} and numerical studies \citep[][]{2025ApJ...983...39R}, large-scale population studies still face major challenges. Existing analytical prescriptions for the secular evolution of mass-transferring systems are either restricted to extremely high eccentricities \citep[][$\delta$-function model]{2007ApJ...667.1170S,2009ApJ...702.1387S} or limited to conservative MT \citep[][emt model]{2019ApJ...872..119H}. It is not trivial to overcome these limitations when performing binary evolution simulations for many stars. For instance, \cite{2025ApJ...983...39R} switch from the $\delta$-function to the classical RLOF model when the eccentricity is $e \leq 0.05$, while \cite{2021MNRAS.502.4479H} present an ad hoc extension of the emt model to the nonconservative regime. These inconsistencies highlight the need for a unified framework that can treat both conservative and nonconservative MT across arbitrary eccentricities, an essential requirement for realistic large-scale binary evolution simulations.

In \citet[][hereafter Paper I]{2025arXiv250905243P}, we developed a unified semianalytic framework for the orbital evolution of mass-transferring binaries. In that work, we treated mass loss, angular momentum loss (AML), and mass exchange as perturbations to the general two-body problem. We demonstrated that in the case of MT via RLOF, additional perturbations arise due to the anisotropic mass ejection from the donor and accretion onto the companion. Consequently, if the AM stored in the binary components is not negligible compared to the orbital AM, then treating the binary components as extended bodies rather than point masses is necessary. For the position of the ejection point, we introduced the global-$L_1$ fit; the most accurate prescription to date for the position of the $L_1$ Lagrangian point as a function of systemic parameters. Additionally, we adopted the phase-dependent MT prescription developed by \cite{2019ApJ...872..119H}, and we derived secular evolution equations for the orbital elements in the adiabatic regime. Finally, we implemented these orbit-averaged equations into the general mass-transfer ({\sc GeMT}) code and applied it to isolated binaries assuming conservative MT (see \citetalias{2025arXiv250905243P} for more detail). Here, we focus on specific scenarios of nonconservative MT.

Nonconservative MT is a process especially relevant for systems containing CO accretors \citep[e.g.,][]{2017A&A...603A.137M,2018MNRAS.480.1580Z}. For COs, accretion is typically limited by radiation pressure in the accretion flow, resulting in very low accretion rates (i.e., Eddington-limited accretion). Additionally, observations confirm that binaries with CO accretors can retain significant eccentricities \citep[e.g.,][]{2005A&AT...24..151R, 2023A&A...671A.149F,2024MNRAS.529.3729S,2024OJAp....7E..58E}. Therefore, modeling orbital evolution under nonconservative MT in eccentric systems is essential for explaining eccentric post-MT binaries, as well as CO mergers detected by LIGO–Virgo–KAGRA \citep[e.g.,][]{2023PhRvX..13a1048A} and future GW sources accessible to LISA \citep[e.g.,][]{2019MNRAS.490.5888L}.

We consider an eccentric binary with component masses $M_{\rm don}$ (donor) and $M_{\rm acc}$ (accretor), defining the mass ratio $q=M_{\rm don}/M_{\rm acc}$.  The orbit has a semimajor axis $a$, an eccentricity $e$, and an orbital period $P_{\rm orb}$. We assume that both stars rotate uniformly at spin angular velocities $\vec{\Omega}_{\rm don}$ and $\vec{\Omega}_{\rm acc}$ aligned with each other and with the orbital angular velocity $\vec{\Omega}_{\rm orb}$. In our semi-detached configuration, the donor loses mass at a rate of $\dot{M}_{\rm don} < 0$ via RLOF, and the accretor gains mass at a rate of $\dot{M}_{\rm acc}$. In the nonconservative regime, a fraction, $\beta$, of the transferred mass is accreted,\footnote{Traditionally, $\beta$ is defined as the fraction of mass transferred from the donor star that is ejected from the system in the vicinity of the accretor \citep[e.g.,][]{1997A&A...327..620S}. Here, we follow the notation introduced in Section 7.2, pages 9–12, of \href{https://www.astro.ru.nl/~onnop/education/binaries_utrecht_notes/Binaries_ch6-8.pdf}{lecture notes on binary star evolution} by Onno Pols.} while the rest escapes the system so that the net mass loss rate is $\dot{M} = (1-\beta) \dot{M}_{\rm don}$. We further parametrize the specific AM carried away by the ejected matter as $\gamma$ (hereafter the ``AML-parameter'') times the specific AM of the binary $J_{\rm orb}/M$:
\begin{equation}\label{eq:angular_momentum_parametrization}
    \frac{\dot{J}_{\rm orb,ml}}{J_{\rm orb}} \equiv \gamma \frac{\dot{M}}{M} = \gamma (1-\beta)\frac{\dot{M}_{\rm don}}{M_{\rm don} + M_{\rm acc}}.
\end{equation}
The AML-parameter is a measure of the efficiency of AM extraction during nonconservative MT.

During RLOF, the donor star transfers mass to its companion, the accretor, through the inner Lagrangian point $L_1$. The position of the $L_1$ point relative to the donor's center of mass is given by the global-$L_1$ fit, $X_{\rm L1}(f_{\rm don},q,e)$, where $f_{\rm don}$ represents the  spin angular velocity of the donor $\Omega_{\rm don}$ normalized to the orbital angular velocity at periapsis (see Appendix B in \citetalias{2025arXiv250905243P}),
\begin{equation}
    f_{\rm don}  \equiv  \frac{\Omega_{\rm don}}{n} \frac{(1-e)^{3/2}}{(1+e)^{1/2}},
\end{equation}
and $n = 2 \pi/P_{\rm orb}$. Similarly, the accretor gains mass from the position $r_{\rm acc}$ relative to its center of mass. This position can be a point on the accretor's surface or the outer edge of the accretion disk.

In \citetalias{2025arXiv250905243P} we made two simplifying assumptions, and we repeat them here for completeness:
\begin{enumerate}
    \item We assumed that any gravitational attractions exerted by the particles in the MT stream on the binary components are negligible.
    \item We assumed that the donor ejects mass with a relative velocity $\vec{\dot{r}}$, that the accretor accretes mass with a relative velocity $-\vec{\dot{r}}$, and that the accretion point corotates with the orbit. 
\end{enumerate}
Under these assumptions, we derived the following orbit-averaged equations of motion for the general case of nonconservative MT via RLOF:
\begin{flalign}
 \frac{\langle \dot{a} \rangle}{a} &= -\frac{2 \langle \dot{M}_{\rm don} \rangle}{M_{\rm don}} \frac{f_{a}(e,x)}{f_{\dot{M}_{\rm don}}(e,x)} \Biggl[ \Biggl(1-\beta q-(1-\beta)\frac{(\gamma+\frac{1}{2})q}{1+q}\Biggr) \nonumber\\
 &+ X_{\rm L1}(f_{\rm don},q,e) g_{a}(e,x) \pm \beta q \frac{r_{\rm acc}}{a} h_{a}(e,x) \Biggr],\label{eq:orbit_averaged_semimajor_axis}\\
 \langle \dot{e} \rangle &= -\frac{2 \langle \dot{M}_{\rm don} \rangle}{M_{\rm don}} \frac{f_{e}(e,x)}{f_{\dot{M}_{\rm don}}(e,x)} \Biggl[ \Biggl(1-\beta q-(1-\beta)\frac{(\gamma+\frac{1}{2})q}{1+q}\Biggr) \nonumber\\
 &+ X_{\rm L1}(f_{\rm don},q,e) g_{e}(e,x) \pm \beta q \frac{r_{\rm acc}}{a} h_{e}(e,x) \Biggr],\label{eq:orbit_averaged_eccentricity}\\
 \langle \dot{\omega} \rangle &= 0, \label{eq:orbit_averaged_argument_of_periapsis}
\end{flalign}
where $x \equiv R_{\rm L}^c/R_{\rm don}$ represents the level at which the physical radius $R_{\rm don}$ overflows the Roche-lobe equivalent radius $R_{\rm L}^c$ \citep{1983ApJ...268..368E} for a circular orbit and $f_{\dot{M}_{\rm don}}(e,x)$, $f_{a}(e,x)$, $f_{e}(e,x)$, $g_{a}(e,x)$, $g_{e}(e,x)$, $h_{a}(e,x)$, and $h_{e}(e,x)$ are dimensionless functions given explicitly in Appendix \ref{app:dimensionless_functions}. The negative sign in front of $r_{\rm acc}$ applies when the ejected mass is accreted on the side of the accretor facing the donor, whereas the positive sign applies when the ejected mass follows a curved trajectory and lands on the far side of the accretor from the donor. Finally, in the limit of point masses, Eqs.~\eqref{eq:orbit_averaged_semimajor_axis} and \eqref{eq:orbit_averaged_eccentricity} simplify to
\begin{flalign}
 \frac{\langle \dot{a} \rangle}{a} &= -\frac{2 \langle \dot{M}_{\rm don} \rangle}{M_{\rm don}} \frac{f_{a}(e,x)}{f_{\dot{M}_{\rm don}}(e,x)} \Biggl[ \Biggl(1-\beta q-(1-\beta)\frac{(\gamma+\frac{1}{2})q}{1+q}\Biggr), 
 \label{eq:orbit_averaged_semimajor_axis_points}\\ \nonumber\\
 \langle \dot{e} \rangle &= -\frac{2 \langle \dot{M}_{\rm don} \rangle}{M_{\rm don}} \frac{f_{e}(e,x)}{f_{\dot{M}_{\rm don}}(e,x)} \Biggl[ \Biggl(1-\beta q-(1-\beta)\frac{(\gamma+\frac{1}{2})q}{1+q}\Biggr). 
 \label{eq:orbit_averaged_eccentricity_points}\\ \nonumber
\end{flalign}

This article is organized as follows. In Sect.~\ref{sec:two}, we introduce various AML scenarios, referring to them as AML modes. Section~\ref{sec:three} presents a parameter exploration of the GeMT model for circular and eccentric orbits, assuming different AML modes. In Sect.~\ref{sec:four}, we compare the parameter space predicted by the GeMT model, under fully nonconservative MT and isotropic reemission, to earlier frameworks. In Sect.~\ref{sec:five}, we apply the GeMT model to isolated binaries undergoing nonconservative MT, and we (1) examine the orbital evolution under different AML modes and (2) compare the predicted evolution with other existing MT frameworks under fully nonconservative MT assuming isotropic reemission for the first time. In Sect.~\ref{sec:six}, we discuss the implications of our work for the evolution of interacting binaries and the predicted population of merging COs. In Sect.~\ref{sec:seven} we summarize and conclude.

\section{Angular momentum loss modes}\label{sec:two}

Specifying the orbital AML-parameter $\gamma$ is not trivial, as it generally varies with orbital phase and depends on the stellar masses, types, and details of the MT process. However, in certain idealized physical scenarios, $\gamma$ can be expressed in terms of known quantities \citep[e.g.,][]{1997A&A...327..620S}. In this work, we explore four different AML modes:
\begin{enumerate}
    \item Jeans (also known as fast wind mode),
    \item Isotropic reemission,
    \item Orbital-AML,
    \item $L_2$ mass loss. 
\end{enumerate}

In the fast wind scenario, mass is lost directly from the donor as a fast, isotropic wind rather than via RLOF. The ejected material carries away the specific orbital AM of the donor, where $D_{\rm don}$ is its distance from the system's center of mass, resulting in an AML rate of $\dot{J}_{\rm orb,ml}/J_{\rm orb} = \Omega_{\rm orb} D_{\rm don}^2$, which corresponds to $\gamma=1/q$. In the classical Jeans approximation, the donor is treated as a point mass (its physical size is neglected), and the wind escapes the system entirely (i.e., $\beta \sim 0$) without interacting with the companion or forming a circumbinary (CB) disk structure. 

\begin{figure}[!htbp]
    \includegraphics[width=\linewidth]{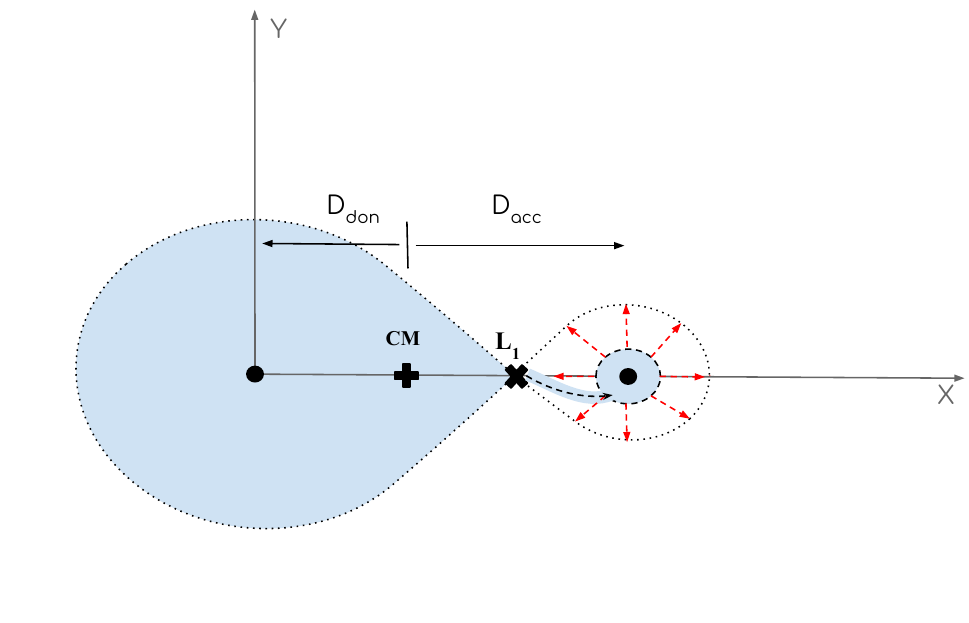}
    \caption{Schematic of mass loss via isotropic reemission. The equipotential surface (dotted curve) is shown in the corotating frame of the binary system. The thick plus sign represents the system's center of mass, and the thick cross represents the location of the $L_1$ point through which the donor transfers mass to the accretor. The mass that is not accreted is ejected isotropically from the vicinity of the accretor as indicated by the red arrows.}
    \label{fig:isotropic_re_emission}
\end{figure}

During isotropic reemission, mass transferred via RLOF cannot be accreted (i.e., $\beta \sim 0$) and is instead ejected isotropically from the vicinity of the accretor. The process is illustrated schematically in Fig.~\ref{fig:isotropic_re_emission}, where $D_{\rm acc}$ denotes the distance of the accretor from the system's center of mass.
In the center-of-mass frame, the ejected mass carries away the specific AM of the accretor, leading to an AML rate given by $\dot{J}_{\rm orb,ml}/J_{\rm orb} = \Omega_{\rm orb} D_{\rm acc}^2$, which corresponds to $\gamma=q$.

In the orbital-AML mode, the ejected mass simply carries the orbital AM. Consequently, the orbital AML-parameter is $\gamma=1$ \cite[e.g.,][]{2008ApJS..174..223B}.

The fourth scenario involves mass loss through the outer Lagrangian point $L_2$.\footnote{In this work, we always refer to the $L_2$ point as the Lagrangian point located behind the accretor, irrespective of whether the accretor is more or less massive than the donor.} In the standard case of RLOF in a circular, synchronous binary, mass transferred from the donor is typically captured by the accretor, either via direct impact or through the formation of an accretion disk. However, \cite{2007ApJ...660.1624S} showed that in eccentric and/or nonsynchronous binaries, the geometry of the quasi-static equipotential surfaces can change such that this is not always the case. Their analysis, focused on the geometry of the equipotential surfaces at periapsis, demonstrated that the quasi-static gravitational potential at $L_2$ can be lower than at $L_1$ across a wide parameter space region (see Appendix~\ref{app:potential_height}). We extend this work by quantifying the impact of $L_2$ mass loss on the orbital evolution during nonconservative MT in eccentric and/or nonsynchronous binaries. Here, we investigate the impact of $L_2$ mass loss on the orbit, assuming stable MT. 

\begin{figure}[!htbp]
    \includegraphics[width=\linewidth]{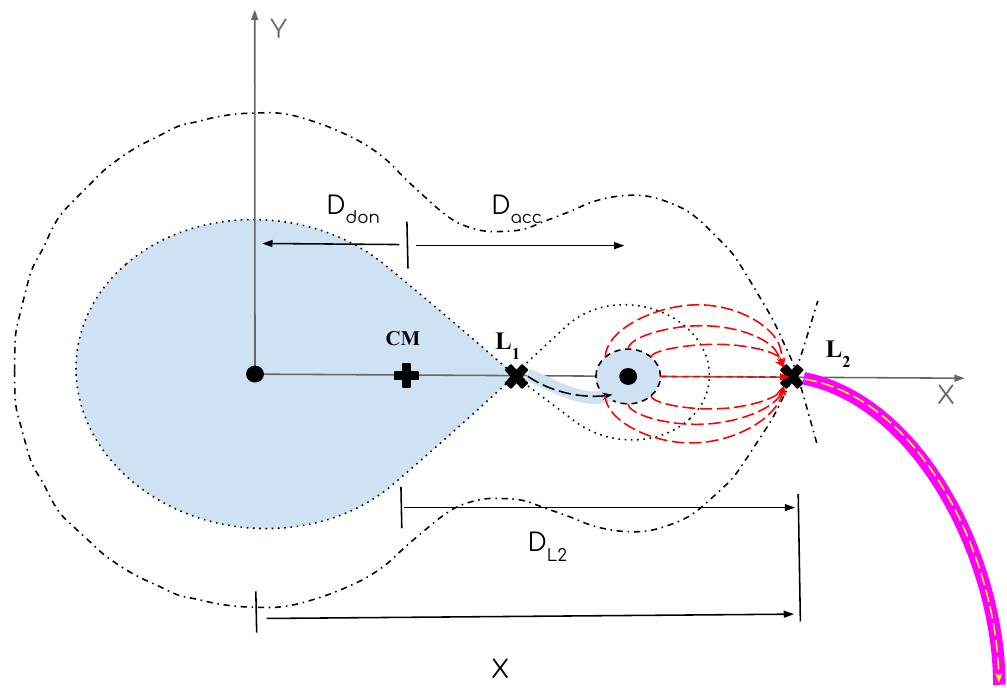}    \caption{Schematic of mass loss through the $L_2$ Lagrangian point. Two equipotential surfaces (dotted and dashed-dotted curves) are shown in the corotating frame of the binary system. The thick plus sign represents the system's center of mass, and the thick crosses represent the locations of the $L_1$, through which the donor transfers mass to the accretor, and $L_2$ points, respectively. The mass that is not accreted is lost through the $L_2$ point indicated by the thick pink line. The red arrows pointing from the accretor to the $L_2$ point are only for schematic purposes.}
    \label{fig:L2_mass_loss}
\end{figure}

We consider MT via RLOF, in which only a fraction of the transferred mass is accreted; the remainder escapes through the $L_2$ Lagrangian point,\footnote{Mass that is lost from $L_2$ can form a CB disk. Interactions between the binary and the CB disk may further affect the orbital evolution \citep[for a comprehensive review see][]{2023ARA&A..61..517L}, but this is beyond the scope of our work.} where $D_{\rm L2}$ is its distance from the system's center of mass (see schematic in Fig.~\ref{fig:L2_mass_loss}). 
We assume this lost mass carries away the specific AM of the $L_2$ point. In the center-of-mass frame, the resulting orbital AML rate satisfies $\dot{J}_{\rm orb,ml}/J_{\rm orb} = \Omega_{\rm orb} D_{\rm L2}^2$, and in Appendix~\ref{app:L_2_position} we demonstrate that the related orbital AML-parameter is
\begin{equation}\label{eq:gamma_L2}
    \gamma= \Bigl(2+q+\frac{1}{q}\Bigr)\Bigl( X_{\rm L2}(f_{\rm don},q,e)-\frac{1}{1+q}\Bigr)^2,
\end{equation}
where $X_{\rm L2}(f_{\rm don},q,e)$, hereafter the ``global-${L_2}$'' fit, gives the position of the $L_2$ point at periapsis in units of the instantaneous distance between the two stars. The global-${L_2}$ fit is given explicitly in Appendix~\ref{app:L_2_position}.

In Fig.~\ref{fig:different_gammas}, we plot $\gamma$ for different AML modes as functions of eccentricity $e$, donor spin $f_{\rm don}$, and mass ratio $q$. For comparison, we also plot $\gamma_{\rm L_2,ml} = 1.44(M_{\rm acc}+M_{\rm don})^2/(M_{\rm acc}M_{\rm don})$ \citep{1998CoSka..28..101P} as used in studies of circular orbits \citep[e.g.,][]{2018ApJ...863....5M,2020ApJ...895...29M,2024MNRAS.532.4826T}. Notably, only the $L_2$ mass-loss mode depends on $e$ and $f_{\rm don}$; the isotropic reemission and Jeans modes depend solely on $q$. 

When the donor is more massive than the accretor ($q>1$), isotropic reemission efficiently extracts orbital AM ($\gamma >1$). This is because the donor orbits closer to the system's center of mass, meaning most of the orbital AM resides in the accretor’s orbit. As expected, the AM carried away by the lost mass decreases as the mass ratio decreases. In contrast, for the Jeans mode, $\gamma$ increases as the accretor becomes more massive than the donor ($q<1$), since most of the orbital AM resides then in the donor’s orbit. When the donor and the accretor have equal masses, then $\gamma = 1$ in both modes.

$L_2$ mass loss consistently yields the highest $\gamma$ across most of the parameter space. This is due to the long lever arm of the $L_2$ point ($D_{\rm L2}$ in Fig.~\ref{fig:L2_mass_loss}) relative to the system's center of mass. As the length of the lever arm changes with orbital parameters, so does the value of $\gamma$ (solid lines in  Fig.~\ref{fig:different_gammas}). Specifically, it increases with decreasing eccentricity across all mass ratios. Moreover, increasingly supersynchronous donors (i.e., higher $f_{\rm don}$) result in lower $\gamma$ values; a trend that is more pronounced for less massive donors ($q < 1$). We conclude that $L_2$ mass loss is by far the most efficient mechanism for extracting orbital AM over a broad range of mass ratios $q$, eccentricities $e$, and donor spin factors $f_{\rm don}$.

\begin{figure}[!htbp]
    \centering
    \includegraphics[width=\linewidth]{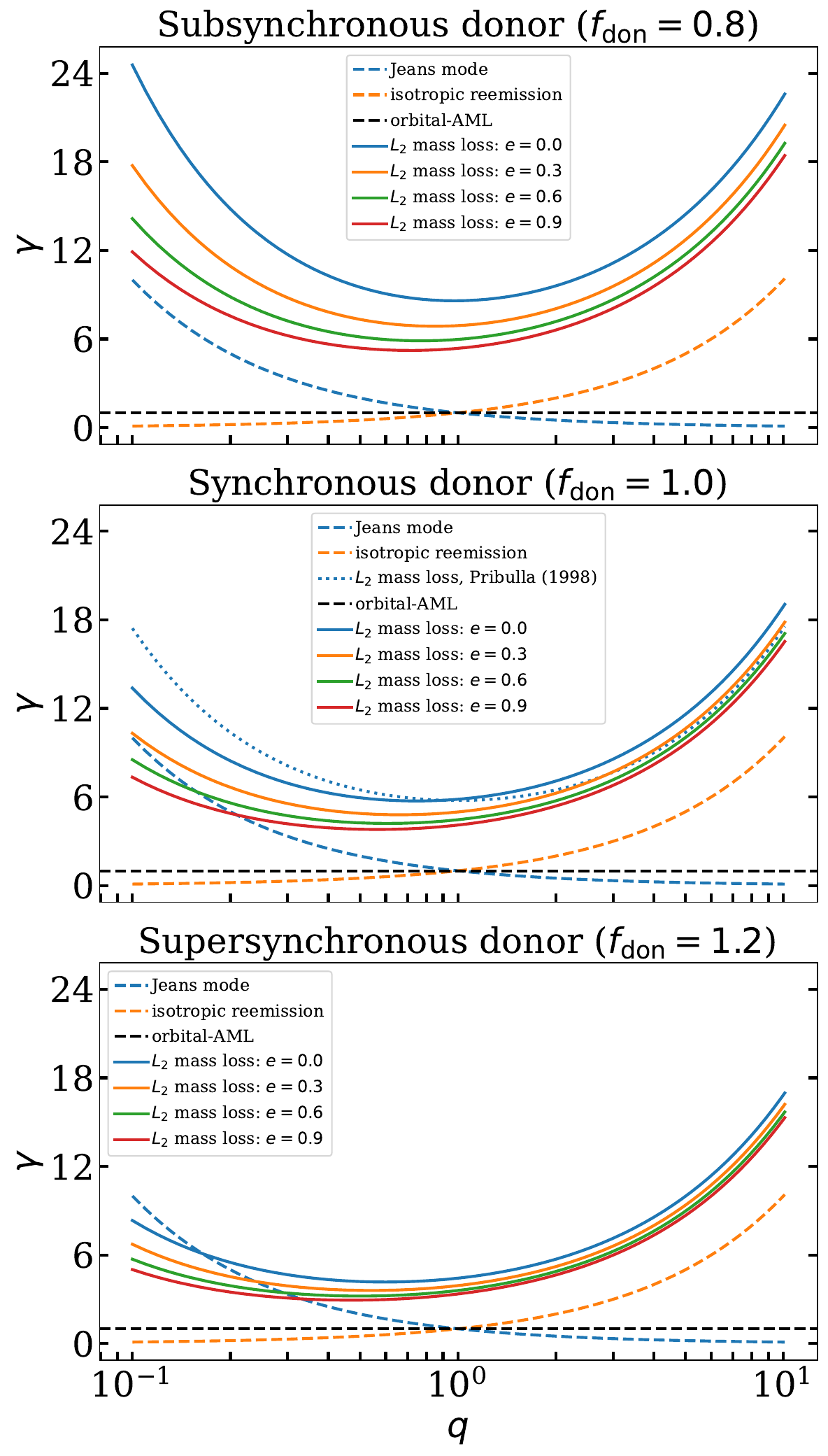}
    \caption{ AML-parameter $\gamma$ for different angular momentum loss modes. From top to bottom, the subfigures correspond to $f_{\rm don}=0.8,1.0,$ and $1.2$, respectively. Dashed blue, orange, and black lines correspond to Jeans, isotropic reemission, and orbital-AML modes, respectively. Solid lines represent the $L_2$ mass-loss mode (Eq.~\ref{eq:gamma_L2}). Blue, orange, green, and red colors correspond to $e=0.0,0.3,0.6,$ and $0.9$, respectively. The dotted blue line corresponds to $L_2$ mass-loss mode for circular orbits  \citep{1998CoSka..28..101P}.}
    \label{fig:different_gammas}
\end{figure}

\begingroup
\begin{table*}[!htbp]
\caption{ Initial conditions at the onset of RLOF.}
  \centering
  \begin{tabular}{cccccccccc}
  \hline
   Fig. & $\beta$ & $\gamma $& $\langle \dot{M}_{\rm don} \rangle$ & $M_{\rm don}$ & $a$ & $e$ & x & $f_{\rm don}$ & $r_{\rm acc}$\\
  &  & & (M$_{\odot} \; \rm yr^{-1}$) & (M$_{\odot}$) & (au) & & & & ($R_{\odot}$) \\
  \hline \hline
   \ref{fig:circ_ang_momentum_loss} & $[0.0,1.0]$ & [q, 1/q, 1, Eq.~\eqref{eq:gamma_L2}] & $ 10^{-4}$ & 26 & 0.16 & 0.0 &–& 1.0 &  0.0  \\
   \ref{fig:semi_major_beta_0.0} & $ 0.0$ & [q, 1/q, 1, Eq.~\eqref{eq:gamma_L2}] & $ 10^{-4}$ & 26 & 0.16 & $ [0.0,0.99]$ & 0.99 & 1.0 &  0.0  \\
   \ref{fig:ecc_beta_0.0} & $0.0$ & [q, 1/q, 1, Eq.~\eqref{eq:gamma_L2}] & $ 10^{-4}$ & 26 & 0.16 & $[0.01,0.99]$ &  0.99 & 1.0 &  0.0  \\
  \hline
  \end{tabular}\label{tab:colormaps_parameters}
  \tablefoot{The spin of the ejection point $f_{\rm don}$ is only relevant to $L_2$ mass loss.}
\end{table*}
\endgroup

\section{Orbital evolution under different AML modes}\label{sec:three}
 
In this section, we investigate the secular evolution of the semimajor axis $a$ and eccentricity $e$ under nonconservative MT. We assume that the AM stored in the stars is negligible compared to the orbital AM, and hence we adopt the point-mass approximation, and the evolution is governed by Eqs.~\eqref{eq:orbit_averaged_semimajor_axis_points} and \eqref{eq:orbit_averaged_eccentricity_points}. The AML-parameter $\gamma$ takes on the values corresponding to the AML modes introduced in Sect.~\ref{sec:two}. Table~\ref{tab:colormaps_parameters} summarizes the relevant parameters for all examples.

\subsection{Circular orbits}\label{subsec:circular_orbits}

\begin{figure}[!htbp]
    \centering
    \includegraphics[width=\linewidth]{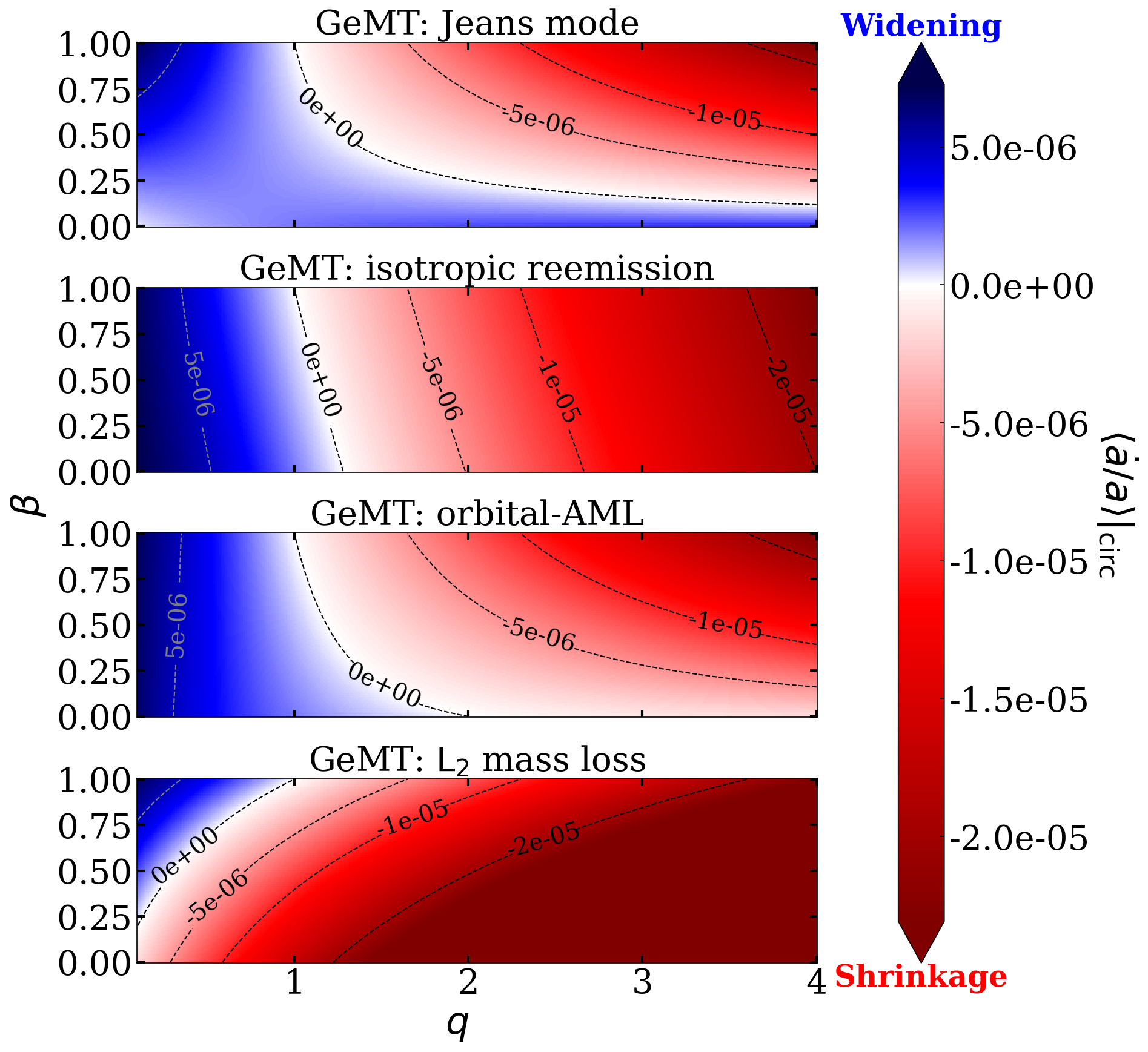}
    \caption{Secular change rate of the semimajor axis as a function of mass ratio $q$ and fraction of accreted mass $\beta$ in the limit of circular orbits for different AML modes. From top to bottom, the GeMT model assuming Jeans mode, isotropic reemission, orbital-AML, and $L_2$ mass loss. The values of the relevant parameters are provided in Table~\ref{tab:colormaps_parameters}.}
    \label{fig:circ_ang_momentum_loss}
\end{figure}

In Figure~\ref{fig:circ_ang_momentum_loss}, we present the secular change rate of the semimajor axis (Eq.~\ref{eq:orbit_averaged_semimajor_axis_points}) as a function of the mass ratio $q$ and the fraction of accreted mass $\beta$, in the limit of circular orbits for different AML modes. Red regions indicate orbital shrinkage, while blue regions indicate orbital widening, with color intensity reflecting the magnitude of the change rate. 

For conservative MT ($\beta = 1$), in the limit of circular orbits and point masses, the transitional mass ratio $q_{\rm trans,a}$ separating orbital widening and shrinkage occurs at equal masses ($q_{\rm trans,a} = 1$). This is shown in Fig.~\ref{fig:circ_ang_momentum_loss}, where all models predict orbital widening for fully conservative MT at $q \leq 1$. This is expected since, for $\beta = 1$, Eqs.~\eqref{eq:orbit_averaged_semimajor_axis_points} and \eqref{eq:orbit_averaged_eccentricity_points} are independent of $\gamma$.

In the general case of nonconservative MT, the orbital response depends strongly on the AML mode. Under Jeans mode, fully nonconservative transfer ($\beta = 0$) always leads to orbital widening (Fig.~\ref{fig:circ_ang_momentum_loss}), but as $\beta$ increases (i.e., more conservative MT), the region of parameter space leading to orbital shrinkage expands; for instance, at $q \approx 4$, the orbit shrinks when $\beta \gtrsim 0.12$. For isotropic reemission and orbital-AML modes, the orbit widens for $q \lesssim 1.28$ and $q \lesssim 2.0$, respectively, under fully nonconservative MT ($\beta=0$). As in Jeans mode, higher $\beta$ (i.e., more conservative MT) reduces the widening region, but both modes predict stronger widening at small $q$, as seen from the deeper blue shades in Fig.~\ref{fig:circ_ang_momentum_loss}. In contrast, $L_2$ mass loss behaves oppositely; the parameter space for orbital shrinkage grows as $\beta$ decreases (i.e., less conservative MT). For fully nonconservative MT, the orbit shrinks for any mass ratio $q$, with stronger shrinkage than in any other AML mode (redder shades in Fig.~\ref{fig:circ_ang_momentum_loss}). In summary, Jeans mode favors widening, while $L_2$ mass loss favors shrinkage, with isotropic reemission and orbital-AML modes in between.

\subsection{Eccentric orbits}\label{subsec:eccentric_orbits}

\begin{figure}[!htbp]
    \centering
    \includegraphics[width=\linewidth]{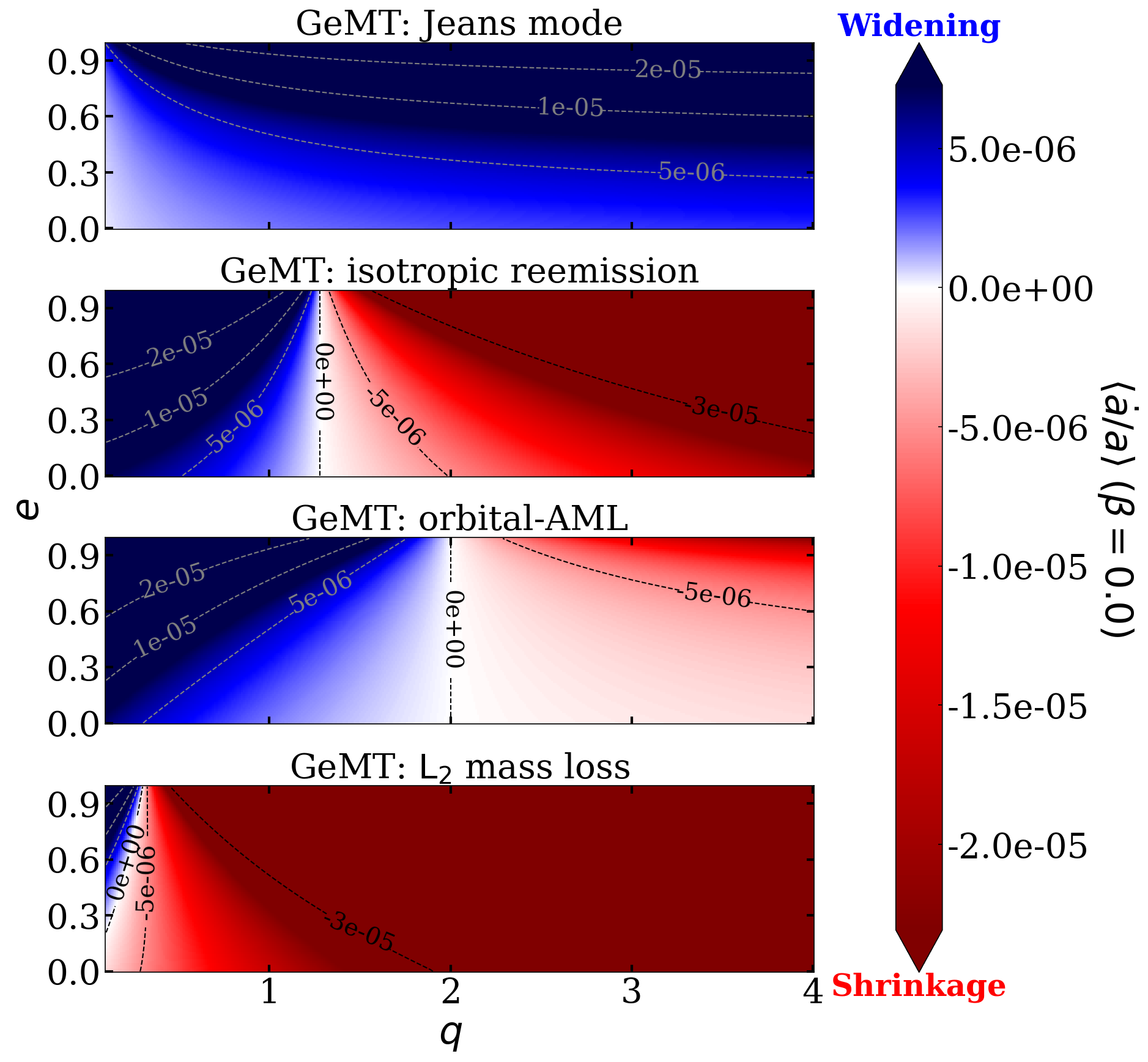}
    \caption{Secular change rate of the semimajor axis as a function of mass ratio $q$ and eccentricity $e$ in the limit of fully nonconservative MT for different AML modes. From top to bottom, the GeMT model assuming Jeans mode, isotropic reemission, orbital-AML, and $L_2$ mass loss. The values of the relevant parameters are provided in Table~\ref{tab:colormaps_parameters}.}
    \label{fig:semi_major_beta_0.0}
\end{figure}

Figure~\ref{fig:semi_major_beta_0.0} shows the secular change rate of the semimajor axis $a$ as a function of mass ratio $q$ and eccentricity $e$ for different AML modes, under fully nonconservative MT ($\beta=0$). This figure is analogous to Fig.~\ref{fig:circ_ang_momentum_loss}, but extends the analysis from circular to eccentric orbits.

During fully nonconservative MT, the Jeans mode yields orbital widening across all mass ratios and eccentricities (Fig.~\ref{fig:semi_major_beta_0.0}). In contrast, isotropic reemission and orbital-AML modes exhibit fixed transitional mass ratios $q_{\rm trans,a} = 1.28$ and $q_{\rm trans,a} = 2.0$, respectively, independent of $e$. When $q>2$, however, the orbital-AML mode drives weaker evolution of the semimajor axis than isotropic reemission for any $e$. The $L_2$ mass-loss mode leads almost exclusively to orbital shrinkage; for $q \gtrsim 0.3$, the orbit always contracts, while widening is restricted to $q \lesssim 0.3$ and $e \gtrsim 0.2$. Across all AML modes, eccentric binaries undergo significantly stronger semimajor axis evolution than circular binaries, as reflected by the color intensity.

\begin{figure}[!htbp]
    \centering
    \includegraphics[width=\linewidth]{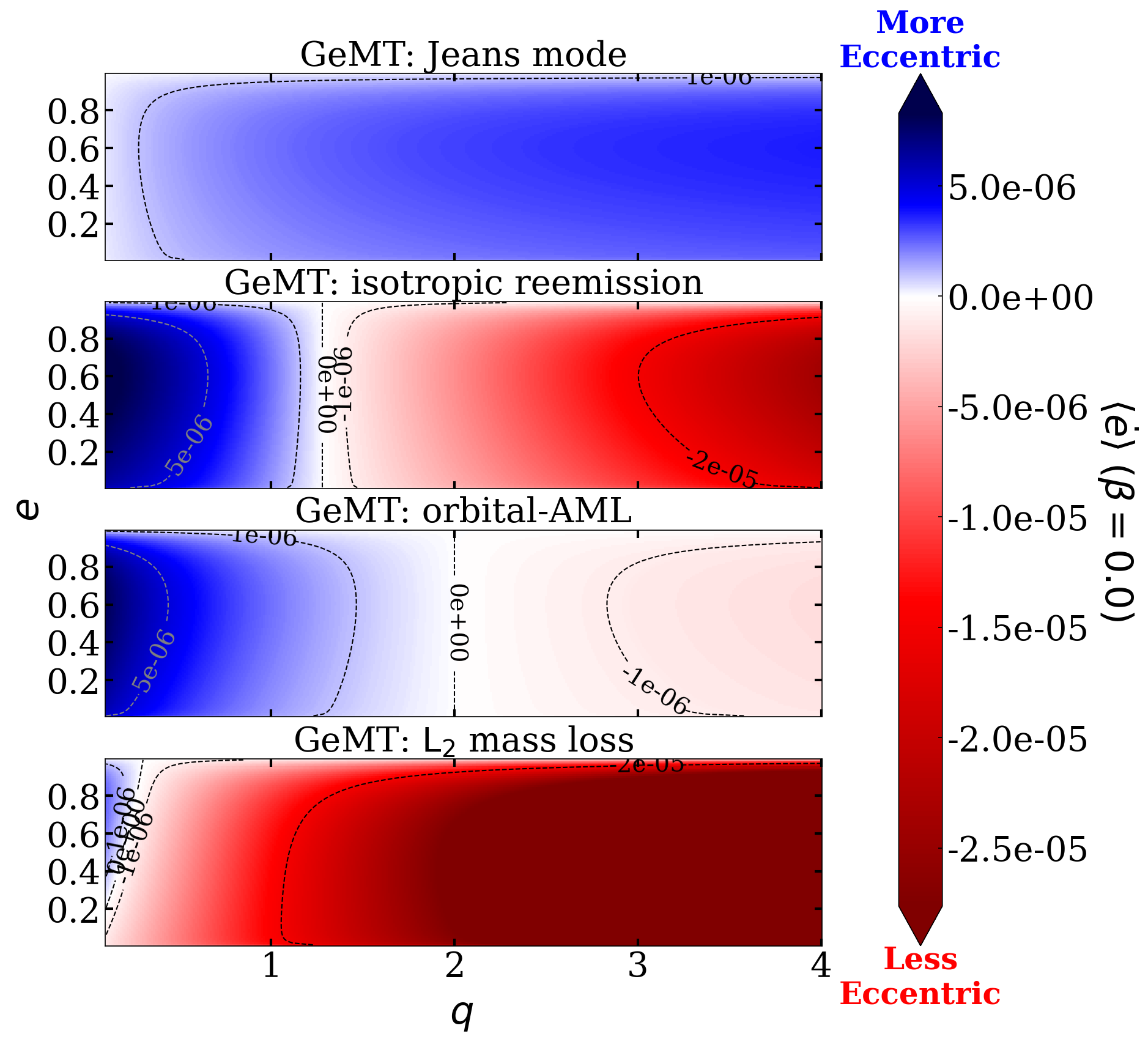}
    \caption{Similar to Fig.~\ref{fig:semi_major_beta_0.0}, but the color gradient now illustrates the secular change rate of the eccentricity.}
    \label{fig:ecc_beta_0.0}
\end{figure}

Figure~\ref{fig:ecc_beta_0.0} is similar to Fig.~\ref{fig:semi_major_beta_0.0}, but now the color gradient indicates the secular change rate of the eccentricity. For a given AML mode, the parameter space for eccentricity evolution resembles that for semimajor axis evolution. Specifically, for the Jeans mode, $\langle \dot{e} \rangle$ is positive across all mass ratios and eccentricities. Isotropic reemission and orbital-AML modes exhibit the same fixed transitional mass ratios $q_{\rm trans,e} = 1.28$ and $q_{\rm trans,e} = 2.0$, respectively, independent of $e$. Furthermore, when $q>2$, the orbital-AML mode drives weaker evolution of the eccentricity than the isotropic reemission for any $e$. The $L_2$ mass-loss mode leads almost exclusively to orbital circularization; for $q \gtrsim 0.3$, the eccentricity always decreases, while eccentricity pumping is restricted to very low mass ratios and $e \gtrsim 0.2$. 

We conclude that binary evolution in eccentric orbits proceeds qualitatively differently from that in circular systems, as phase-dependent RLOF inherently drives eccentricity evolution. Quantitatively, higher $e$ values yield stronger $\langle \dot{a} \rangle$ at fixed $q$ (see color intensity in Figs.~\ref{fig:circ_ang_momentum_loss} and \ref{fig:semi_major_beta_0.0}). For a given AML mode and mass-transfer efficiency $\beta$, the secular changes in $a$ and $e$ are correlated: the orbit either widens while becoming more eccentric, or shrinks while circularizing (Figs.~\ref{fig:semi_major_beta_0.0} and \ref{fig:ecc_beta_0.0}).

For both circular and eccentric orbits, Jeans mode generally drives orbital widening and eccentricity pumping, while $L_2$ mass loss drives orbital shrinkage and circularization, with isotropic reemission and orbital-AML modes producing intermediate behavior. For modes 1 to 3, increasingly nonconservative MT (i.e., lower $\beta$) enlarges the parameter space for both orbital widening and eccentricity pumping (i.e., larger $q_{\rm trans,a}$ and $q_{\rm trans,e}$). In contrast, for mode 4, decreasing $\beta$ has the opposite effect: shrinking this parameter space and yielding smaller $q_{\rm trans,a}$ and $q_{\rm trans,e}$ (see Fig.~\ref{fig:circ_ang_momentum_loss}).\footnote{Figure~\ref{fig:circ_ang_momentum_loss} depicts circular orbits, yet we verified that the aforementioned trend holds for any $e$.}

\section{Comparison to earlier mass-transfer frameworks}\label{sec:four}

In this section, we systematically compare the GeMT model to the $\delta$-function \citep{2009ApJ...702.1387S} and the emt \citep{2021MNRAS.502.4479H} models; to our knowledge, these three frameworks are the only ones that describe secular orbital evolution due to nonconservative MT in eccentric binaries. We adopt isotropic reemission as the fiducial orbital AML mode for this comparison. However, we emphasize that the ad hoc extension of the emt model to the nonconservative case does not account for the effects of AML on the evolution of the orbital elements; in this model, the orbital evolution is independent of the AML-parameter $\gamma$, and thus we cannot adopt any specific AML mode. Moreover, the $\delta$-function model does not conserve orbital AM in the limit of conservative MT (see Sect. 7.3 and Appendix D in \citetalias{2025arXiv250905243P}), yet since these equations have been used in the literature \citep[e.g.,][]{2007ApJ...667.1170S,2009ApJ...702.1387S,2025ApJ...983...39R} as such, we apply them here without the proper normalization factor. We organize the comparison into circular and eccentric cases in Sects.~\ref{subsec:circular_orbits_2} and \ref{subsec:eccentric_orbits_2}, respectively. 

For circular orbits, the GeMT (in the point-mass limit; Eq.~\ref{eq:orbit_averaged_semimajor_axis_points}) and the $\delta$-function models both reduce to the canonical relation for the evolution of the semimajor axis, with the latter predicting weaker evolution of $a$ (see also Sect. 5.1 in \citetalias{2025arXiv250905243P}).\footnote{The $\delta$-function model is not valid at low eccentricities. For instance, at $e=0.0$ it predicts a nonzero eccentricity derivative, becoming negative when $q>1$ and positive when $q<1$, as shown by \cite{2019ApJ...872..119H}.} Thus, we compare GeMT-model predictions for extended bodies (i.e., Eqs.~\ref{eq:orbit_averaged_semimajor_axis} and \ref{eq:orbit_averaged_eccentricity}) to these in the point-mass limit (i.e., Eqs.~\ref{eq:orbit_averaged_semimajor_axis_points} and \ref{eq:orbit_averaged_eccentricity_points}) and to the ad hoc extension of the emt model. For eccentric orbits, we also include the $\delta$-function model. In the general case of extended bodies, the ejection point is taken to be the $L_1$ Lagrangian point. For the $\delta$-function and emt models, we adopt the prescriptions derived by \cite{2007ApJ...667.1170S} (Eq. A15 in their Appendix A) and \cite{2019ApJ...872..119H} (Eq. 40 in their Appendix A; hereafter ``low-$f_{\rm don}$'' model), respectively.  Table~\ref{tab:colormaps_parameters_2} summarizes the relevant parameters for all examples.

\subsection{Circular orbits}\label{subsec:circular_orbits_2}

\begin{figure}[!htbp]
    \centering
    \includegraphics[width=\linewidth]{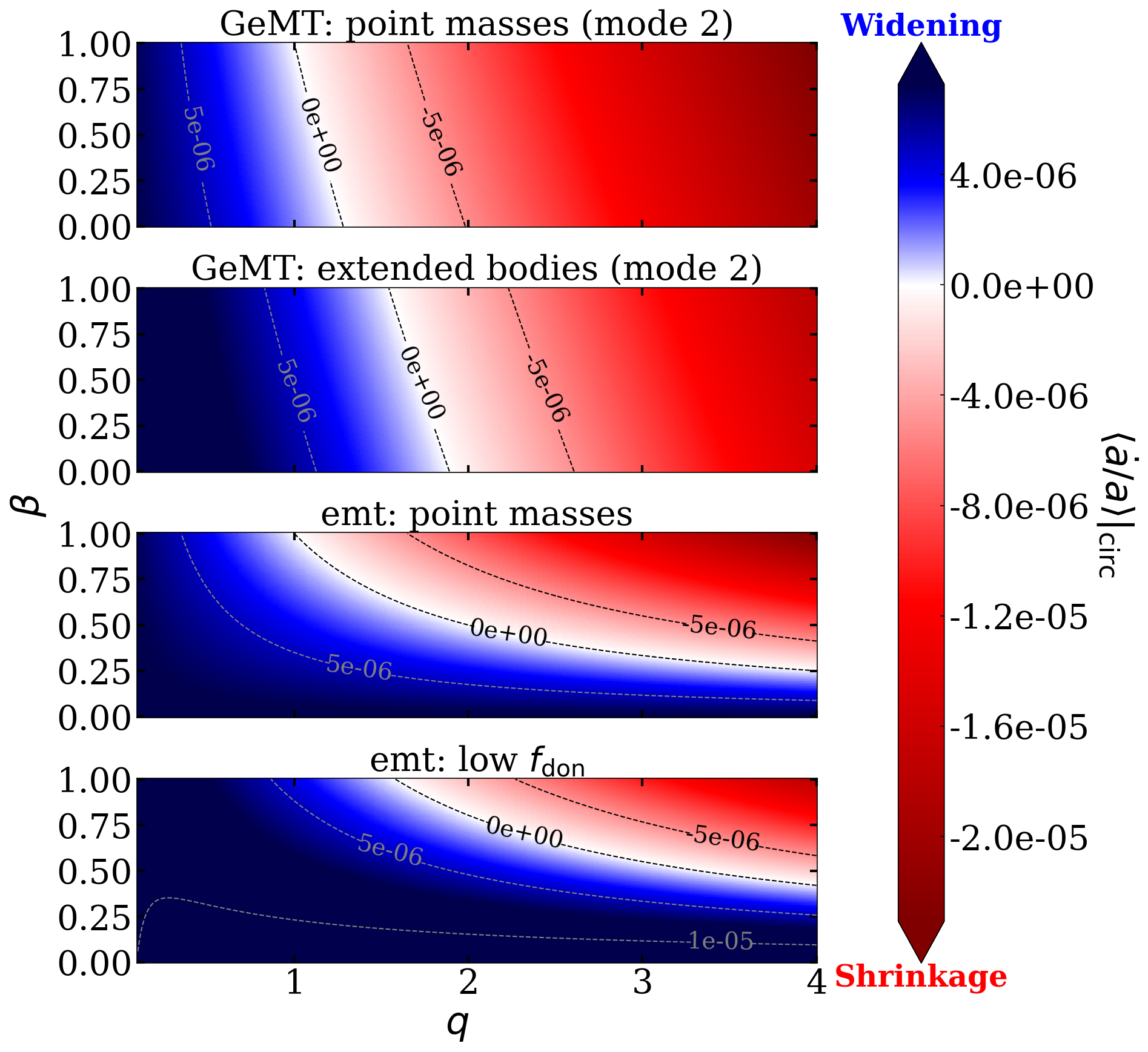}
    \caption{Secular change rate of the semimajor axis as a function of mass ratio $q$ and fraction of accreted mass $\beta$ in the limit of circular orbits assuming isotropic reemission. From top to bottom, the GeMT model in the limit of point masses (Eq.~\ref{eq:orbit_averaged_semimajor_axis_points}), extended bodies (Eq.~\ref{eq:orbit_averaged_semimajor_axis}), the emt model in the limit of point masses and extended bodies. The values of the relevant parameters are provided in Table~\ref{tab:colormaps_parameters}.}
    \label{fig:colormap_semimajor_axis_circ}
\end{figure}

In Figure~\ref{fig:colormap_semimajor_axis_circ}, we plot the secular change rate of the semimajor axis as functions of mass ratio $q$ and accretion efficiency $\beta$, in the limit of circular orbits. In the general case of extended bodies (Eq.~\ref{eq:orbit_averaged_semimajor_axis}), the GeMT model accounts for the effects of anisotropic ejection and accretion on the orbital evolution, resulting in deviations from the classical RLOF picture. Specifically, both the parameter space for orbital widening and the intensity of the blue region increase (Fig.~\ref{fig:colormap_semimajor_axis_circ}). As a result, $q_{\rm trans,a}$ shifts to higher values (e.g., from $q_{\rm trans,a} = 1.28$ to $q_{\rm trans,a} \approx 1.89$ for $\beta = 0$), while the bluer regions in the figure indicate stronger orbital widening compared to the point-mass limit. 

The emt model also predicts $q_{\rm trans,a} > 1$, but the parameter space differs significantly from that predicted by the GeMT model (Fig.~\ref{fig:colormap_semimajor_axis_circ}). In the limit of point masses, the orbit widens for all mass ratios $q$ under fully nonconservative MT ($\beta = 0$). As $\beta$ increases (i.e., more conservative MT), however, the region of parameter space leading to orbital shrinkage grows; for instance, at $q \approx 4$, the orbit shrinks for $\beta \gtrsim 0.25$. Moreover, the intensity of the blue region indicates stronger orbital widening compared to the GeMT model. Finally, when accounting for extended bodies (low-$f_{\rm don}$ model in Fig.~\ref{fig:colormap_semimajor_axis_circ}), both the parameter space for orbital widening and the intensity of the blue region increase similar to the GeMT model. 

\begingroup
\begin{table*}[!htbp]
\caption{ Initial conditions at the onset of RLOF.}
  \centering
  \begin{tabular}{cccccccccc}
  \hline
  Fig. & $\beta$ & $\gamma $& $\langle \dot{M}_{\rm don} \rangle$ & $M_{\rm don}$ & $a$ & $e$ & x & $f_{\rm don}$ & $r_{\rm acc}$\\
  &  & & (M$_{\odot} \; \rm yr^{-1}$) & (M$_{\odot}$) & (au) & & & & ($R_{\odot}$) \\
  \hline \hline
  \ref{fig:colormap_semimajor_axis_circ} & $[0.0,1.0]$ & q & $ 10^{-4}$ & 26 & 0.16 & 0.0 &–& 1.0 &  0.0  \\
  \ref{fig:colormap_semimajor_axis_ecc} & $0.0$ & q & $ 10^{-4}$ & 26 & 0.16 & $[0.0,0.99]$ & 0.99 & 1.0 &  0.0  \\
  \ref{fig:colormap_eccentricity_ecc} & $0.0$ & q & $ 10^{-4}$ & 26 & 0.16 & $[0.01,0.99]$ &  0.99 & 1.0 &  0.0  \\
  \hline
  \end{tabular}\label{tab:colormaps_parameters_2}
  \tablefoot{The $\langle \dot{M}_{\rm don} \rangle $ term is not applicable to the $\delta$-function model, instead an instantenaneous mass-transfer rate $\dot{M}_{0} = 10^{-4}$ M$_{\odot} \; \rm yr^{-1}$ is used. The spin of the ejection point $f_{\rm don}$ is only applicable to the $\delta$-function and GeMT models in the general case of extended bodies. The parameter $x$ is only applicable to the emt and GeMT models. The emt model does not consider different angular momentum loss modes; it is independent of the AML-parameter $\gamma$.}
\end{table*}
\endgroup

\subsection{Eccentric orbits}\label{subsec:eccentric_orbits_2}

\begin{figure}[!htbp]
    \centering
    \includegraphics[width=\linewidth]{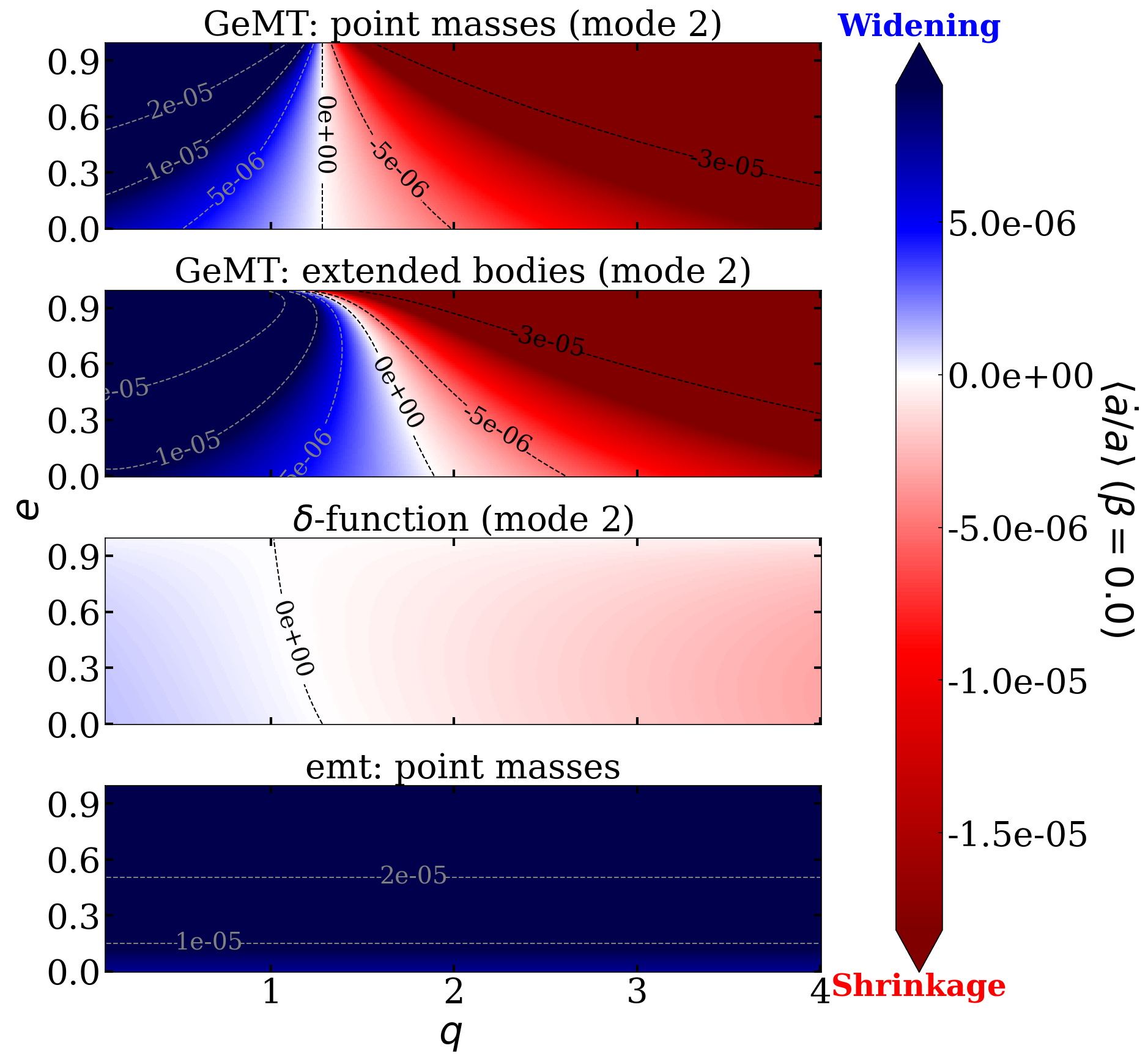}
    \caption{Secular change rate of the semimajor axis as a function of mass ratio $q$ and eccentricity $e$ under fully nonconservative MT. From top to bottom, the GeMT model in the point-mass limit, the $\delta$-function model, the emt model, and the GeMT model for extended bodies. The values of the relevant parameters are listed in Table~\ref{tab:colormaps_parameters}.}
    \label{fig:colormap_semimajor_axis_ecc}
\end{figure}

In Figure~\ref{fig:colormap_semimajor_axis_ecc}, we present the secular change rate of the semimajor axis $a$ as functions of mass ratio $q$ and eccentricity $e$, assuming fully nonconservative MT ($\beta = 0$). We also compare these results with the secular rates predicted by the GeMT model in the point-mass limit, as well as with those from the $\delta$-function and emt models.

For the GeMT model, in the limit of point masses, the transitional mass ratio is $q_{\rm trans,a} = 1.28$, independent of the eccentricity. When accounting for extended bodies, the parameter space for orbital widening increases, and $q_{\rm trans,a}$ depends on $e$, however. Specifically, for decreasing eccentricity, $q_{\rm trans,a}$ shifts to higher values, and conversely, at higher eccentricity, it decreases. The $\delta$-function model also predicts a decrease in $q_{\rm trans,a}$ with increasing $e$; its predicted $q_{\rm trans,a}$ values are significantly lower, however. At $e=0.99$ and $e=0.0$, the orbit widens for any $q \leq 1.0$ and $q \leq 1.28$, respectively (Fig.~\ref{fig:colormap_semimajor_axis_ecc}). Additionally, the color gradient in the $\delta$-function model indicates a weaker evolution of $a$ compared to the GeMT model. Finally, the emt model predicts orbital widening for any $q$ and $e$, with stronger widening at higher eccentricities.

\begin{figure}[!htbp]
    \centering
    \includegraphics[width=\linewidth]{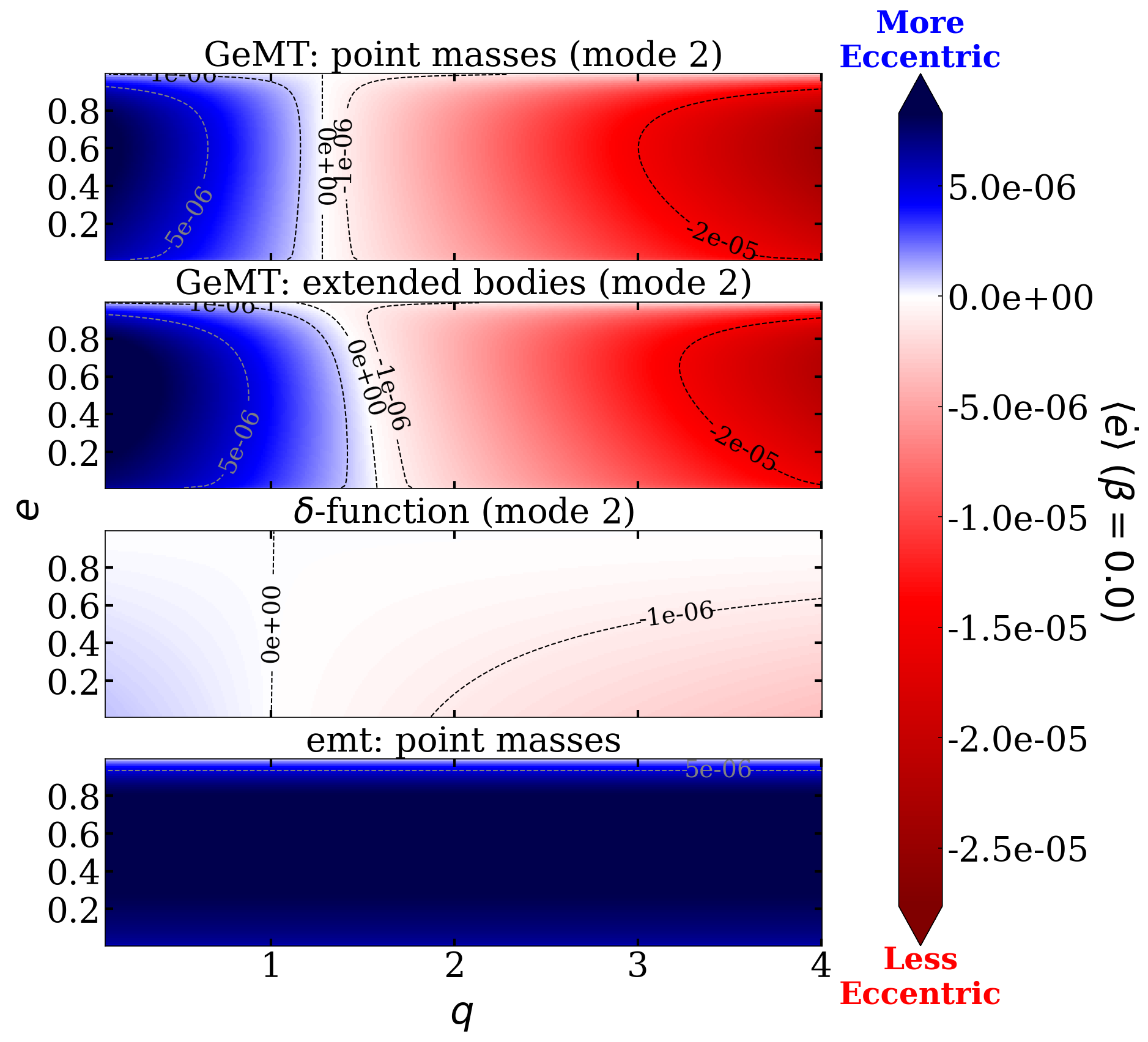}
    \caption{Similar to Fig.~\ref{fig:colormap_semimajor_axis_ecc}, but the color gradient now illustrates the secular change rate of the eccentricity.} 
    \label{fig:colormap_eccentricity_ecc}
\end{figure}

Figure~\ref{fig:colormap_eccentricity_ecc} is similar to Fig.~\ref{fig:colormap_semimajor_axis_ecc}, but the color gradient now illustrates the secular change rate of the eccentricity. In the limit of point masses, the GeMT model predicts that $\langle \dot{e} \rangle$ is positive for $q<1.28$ and negative for $q>1.28$, independent of $e$. When extended bodies are considered, the GeMT model yields a broader parameter space for eccentricity pumping; the transitional mass ratio, $q_{\rm trans,e}$ shifts to higher values, a trend more prominent for low eccentricities. Specifically, at $e=0.99$ and $e=0.01$, the eccentricity increases for any $q \leq 1.12$ and $q \leq 1.59$, respectively. In contrast, the $\delta$-function model maintains $q_{\rm trans,e}$ nearly independent of $e$, restricting the eccentricity-pumping regime to $q \leq 1.0$. Additionally, the color gradient for this model indicates a weaker evolution of eccentricity compared to the GeMT model. Finally, the emt model predicts the strongest evolution of the eccentricity, which increases for any mass ratio $q$ and $e \geq 0.01$.

This comparison highlights that predictions for nonconservative MT differ substantially between different frameworks. Overall, the emt model yields the strongest evolution of both the semimajor axis and eccentricity, typically favoring orbital widening and eccentricity growth, even when the donor is much more massive than the accretor ($q > 1$). By contrast, the $\delta$-function model predicts the weakest orbital evolution, with widening and eccentricity growth largely confined to small mass ratios ($q \lesssim 1$). Finally, the GeMT model yields intermediate behavior.

Figures~\ref{fig:colormap_semimajor_axis_ecc} and \ref{fig:colormap_eccentricity_ecc} show that when accounting for anisotropic ejection (i.e., extended bodies; $X_{\rm L1}(f_{\rm don},q,e) \neq 0$) the parameter space differs. The spin of the donor $f_{\rm don}$ further modifies these domains by shifting the position of the instantaneous $L_1$ point in the corotating frame. Although we do not explore this effect in detail here (see ), the trend is the same in both circular and eccentric orbits: subsynchronous donors ($f_{\rm don} < 1$) expand the domains for orbital widening and eccentricity pumping (higher $q_{\rm trans,a}$ and $q_{\rm trans,e}$), while supersynchronous donors (higher $f_{\rm don} > 1$) contract them (lower $q_{\rm trans,a}$ and $q_{\rm trans,e}$). This dependence on $f_{\rm don}$ becomes less pronounced at higher eccentricities.

\section{Applications. Nonconservative mass transfer}\label{sec:five}

In this section, we examine the orbital evolution of isolated binary systems undergoing nonconservative RLOF by performing numerical integrations. Here, we neglect additional physical processes such as tides or the evolution of the stellar components in the binary to isolate the effects of nonconservative MT via RLOF.  We treat the stars as rigid spheres and assume that both the donor's radius $R_{\rm don}$ and the accretor's radius (or the outer edge of the accreting disc) $r_{\rm acc}$ remain constant throughout the integration. 

We consider a system in which a hydrogen-rich donor initiates RLOF onto a black hole (BH) accretor during the early Hertzsprung gap (HG). The initial parameters are $M_{\rm don} = 26$ M$_{\odot}$, $M_{\rm acc} = 10$ M$_{\odot}$, $a = 0.16$ au, and $e=0.25$. We further assume that the donor is synchronous with the orbital angular velocity at periapsis, such that $f_{\rm don}=1$, and that initiates RLOF at $R_{\rm don} \approx 13$ R$_{\odot}$, corresponding to $x \approx 1.23$. Accretion onto BHs is typically Eddington-limited, and we therefore adopt $\beta = 0$. Moreover, we assume that $r_{\rm acc} = 0$ R$_{\odot}$. In addition, we choose $\dot{M}_{0} = 10^{-4}$ M$_\odot$ yr$^{-1}$ for the $\delta$-function model. We choose an orbit-averaged mass-transfer rate of $\langle \dot{M}_{\rm don} \rangle = 10^{-4}$ M$_{\odot}$ yr$^{-1}$ for the emt and GeMT models, although the qualitative evolution of the orbital parameters is largely independent of this specific value. Hereafter, this system serves as our fiducial model for comparison.

In Section~\ref{subsec:AML_modes}, we compare the orbital evolution predicted by the GeMT model under the AML modes described in Sect.~\ref{sec:two}. In Section~\ref{subsec:models_comparison}, we further compare the GeMT-model predictions with those of earlier MT frameworks. In all cases, we evolve the system until the donor star loses its envelope, leaving behind a naked helium core with mass $M_{\rm don,core} = 9$ M$_{\odot}$, corresponding to $f_{\rm core} = M_{\rm don,core} / M_{\rm don,ZAMS} \approx 0.34$  \citep{1998MNRAS.298..525P,2022ApJ...940..184V}. Moreover, we track the orbital evolution assuming point masses under the assumption that the AM stored in the binary components is negligible compared to the orbital AM.\footnote{In contrast to the GeMT and emt models, the $\delta$-function model is derived mathematically \citep[see][]{2007ApJ...667.1170S,2009ApJ...702.1387S}. Hence, we account for the ejection point, which is included in the derivation, via the prescription derived by \cite{2007ApJ...667.1170S} (Eq. A15 in their Appendix A).}

\subsection{GeMT-model applications under different AML modes}\label{subsec:AML_modes}

\begin{figure}[!htbp]
    \centering
    \includegraphics[width=\linewidth]{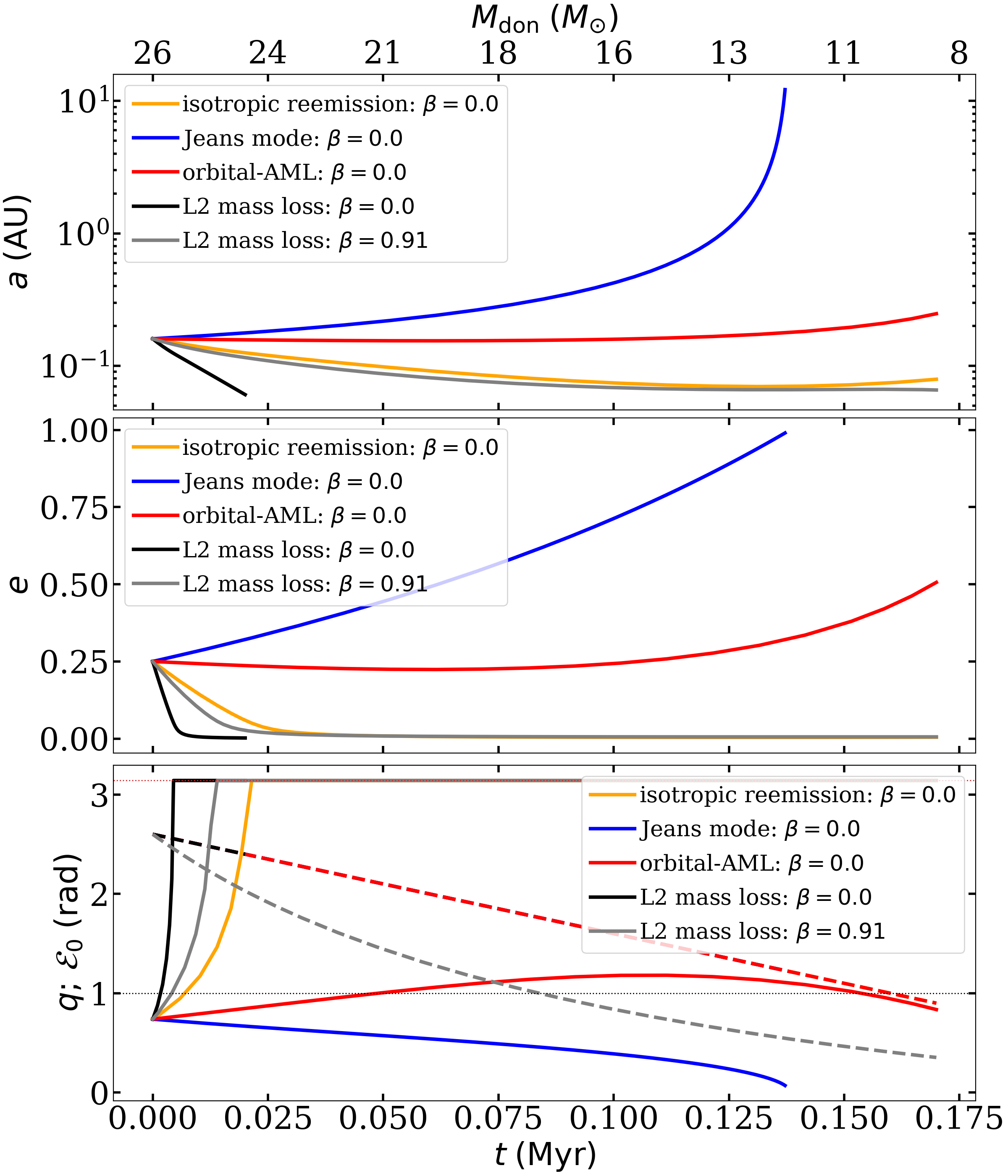}
    \caption{Evolution of the semimajor axis (top), eccentricity (middle), and both the mass ratio $q$ and angle $\mathcal{E}$ (bottom) as a function of time for different AML modes for our example model of a 26 M$_{\odot}$ donor with a 10 M$_{\odot}$ accreting BH companion (point-mass limit). The orange, blue, and red lines correspond to the isotropic reemission, Jeans, and orbital-AML modes, respectively. The black and gray lines represent the $L_2$ mass-loss mode for $\beta = 0$ (fully nonconservative MT) and $\beta=0.91$ (highly conservative MT), respectively. For all models $f_{\rm don}=1$. In the bottom subfigure, the two horizontal dotted lines indicate $\mathcal{E}_0 = \pi$ and $q = 1$.}
    \label{fig:compare_AML_modes}
\end{figure}

\begin{figure}[!htbp]
    \centering
    \includegraphics[width=\linewidth]{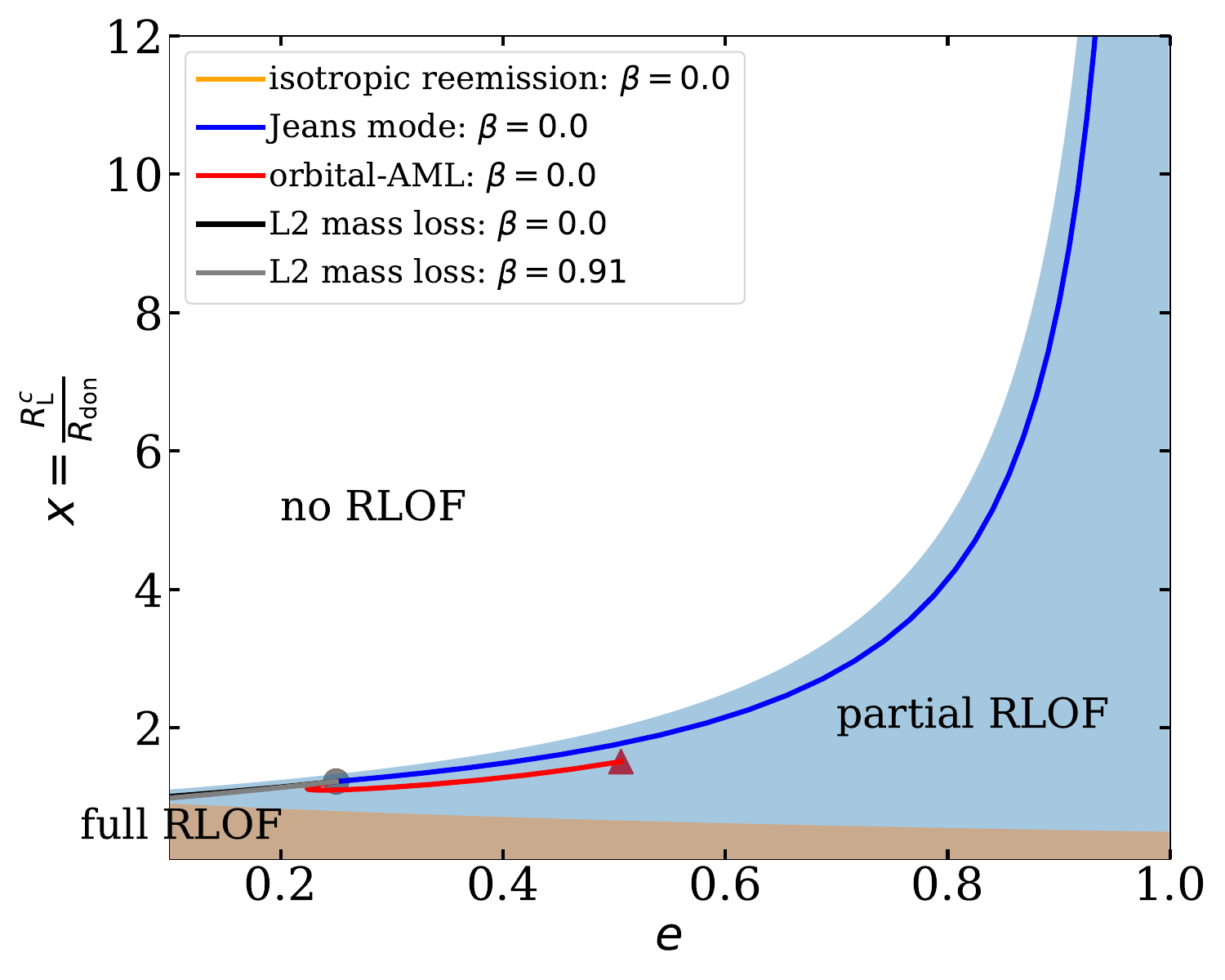}
    \caption{Graphical representation of the mass-transfer regimes. The white region indicates no mass transfer, i.e., the system is detached. The light blue region corresponds to partial RLOF, where mass transfer occurs during part of the orbit. The light-brown region represents full RLOF, with continuous mass transfer throughout the entire orbit. The orange, blue, and red lines correspond to the isotropic reemission, Jeans, and orbital-AML modes, respectively. The black and gray lines represent the $L_2$ mass-loss mode for $\beta = 0$ (fully nonconservative MT) and $\beta=0.91$ (highly conservative MT), respectively. The circles and triangles indicate the initial and final positions of the systems, respectively. The final position for the blue line is $x \approx 80$, but we limited the y-axis for visualization purposes.}
    \label{fig:eccentricity_x_GeMT_AML_modes}
\end{figure}

In Figure~\ref{fig:compare_AML_modes} we present the system's evolution predicted by the GeMT model assuming Jeans, isotropic reemission, orbital-AML, and $L_2$ mass-loss modes. Moreover, Fig.~\ref{fig:eccentricity_x_GeMT_AML_modes} illustrates MT evolves with the orbit under the aforementioned AML modes; the system transitions between partial and full RLOF. Initially, all systems experience partial RLOF. 

For the Jeans mode (blue lines), the orbit widens and becomes more eccentric. As the eccentricity increases, MT becomes increasingly constrained around periastron; this can be seen both at the bottom of Fig.~\ref{fig:compare_AML_modes}, where $\mathcal{E}_0 \rightarrow 0$, and at Fig.~\ref{fig:eccentricity_x_GeMT_AML_modes} where the system asymptotically approaches the no RLOF regime (see also Appendix D in \citetalias{2025arXiv250905243P}). The evolution proceeds until $t \approx 0.137$ Myr; at that point the orbit becomes parabolic (i.e., $e \geq 0.99$) and the integration terminates. 

In contrast, for the orbital-AML mode (red lines), the orbit shrinks at the onset of RLOF, and the eccentricity decreases. At $t \approx 0.051$ Myr, however, when $M_{\rm don} \leq 20$ M$_{\odot}$, the system enters the orbital widening and eccentricity pumping regime (see also Figs.~\ref{fig:semi_major_beta_0.0} and \ref{fig:ecc_beta_0.0}). Throughout its evolution, the system experiences partial RLOF (i.e., $0 < \mathcal{E}_0 < \pi$; bottom of Fig.~\ref{fig:compare_AML_modes} and Fig.~\ref{fig:eccentricity_x_GeMT_AML_modes}), which results in a post-MT orbit with $a \approx 0.25$ au, $e \approx 0.51$, $M_{\rm don} = 9$ M$_{\odot}$, and $M_{\rm acc} = 10$ M$_{\odot}$.

For the isotropic reemission (orange lines) and $L_2$ mass-loss (black lines), the system experiences strong orbital shrinkage and eccentricity damping at the onset of RLOF. The system transitions from partial to full RLOF (i.e., $\mathcal{E}_0 = \pi$; see bottom panel of Fig.~\ref{fig:compare_AML_modes} and Fig.~\ref{fig:eccentricity_x_GeMT_AML_modes}) at $t \approx 0.021$ Myr for isotropic reemission and $t \approx 0.005$ Myr for $L_2$ mass loss (black line).  After this transition, the eccentricity remains small but nonzero. Notably, $L_2$ mass loss leads to a stronger decrease in the semimajor axis and eccentricity, ultimately causing the system to merge at $t \approx 0.02$ Myr (black line). In contrast, isotropic reemission yields a post-MT system with $a \approx 0.08$ au, $e \approx 0.005$, $M_{\rm don} = 9$ M$_{\odot}$, and $M_{\rm acc} = 10$ M$_{\odot}$. 

\begin{figure}[!htbp]
    \centering
    \includegraphics[width=\linewidth]{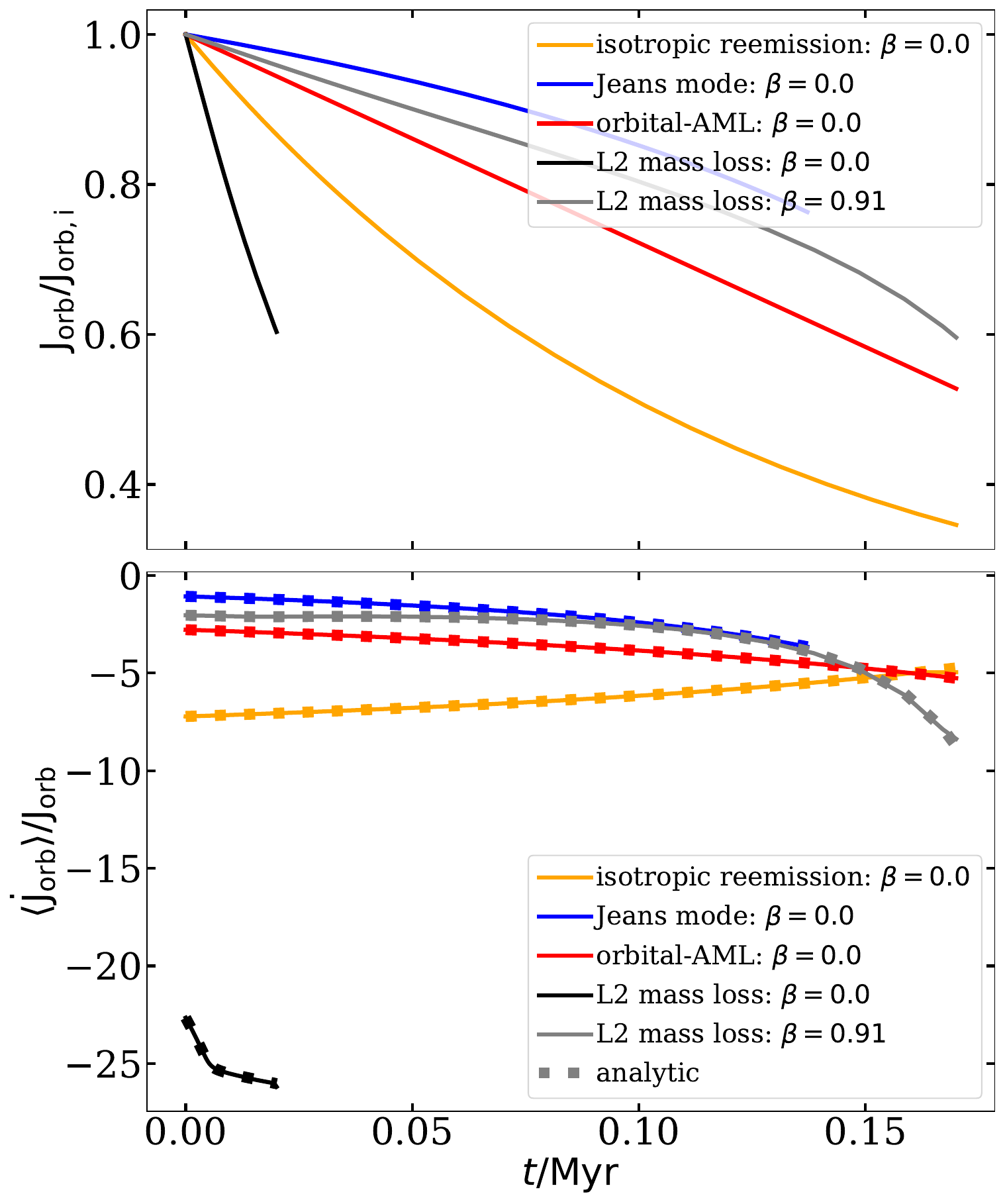}
    \caption{Evolution of the orbital angular momentum (top) and its secular change rate  (bottom) for the examples presented in Fig.~\ref{fig:compare_AML_modes}. The solid lines correspond to the numerical solutions, and the square markers to the analytical expectations given by orbit averaging Eq.~\eqref{eq:angular_momentum_parametrization}. The AML-parameter is $\gamma = q, 1/q,$ and $ 1.0$, for isotropic reemission, Jeans, and orbital-AML modes, respectively. For the $L_2$ mass-loss mode, $\gamma$ is given by Eq.~\eqref{eq:gamma_L2}.}
    \label{fig:angular_momentum}
\end{figure}

In the top panel of Fig.~\ref{fig:angular_momentum}  we present the numerically calculated evolution of the orbital AM under the aforementioned AML modes. In the bottom panel, we compare the numerically calculated secular change rate of the orbital AM to the analytical expectations from orbit averaging Eq.~\eqref{eq:angular_momentum_parametrization} to ensure consistency. For isotropic reemission, Jeans, and orbital-AML modes, $\gamma = q, 1/q,$ and $1.0$, respectively. For the $L_2$ mass-loss mode $\gamma$ is given by Eq.~\eqref{eq:gamma_L2}.
It is clear that the numerical results are in excellent agreement with the analytical predictions. 

In summary, the orbital evolution and the final fate of mass-transferring systems are strongly dependent on the adopted AML mode. $L_2$ mass loss is by far the most efficient mechanism for extracting orbital AM. To demonstrate the effect of $L_2$ mass loss on the orbital evolution, we include a model in which less than 10\% of the transferred mass is lost via $L_2$, while the remaining is accreted by the companion (gray line in Fig.~\ref{fig:angular_momentum}). Notably, even in this fairly conservative MT scenario, the orbit still shrinks, resulting in a compact post-MT binary with $a \approx 0.06 $ au and $e \approx 0.005$. We note that $\beta = 0.91$ is not a realistic value for a BH accretor, and it is chosen here for illustrative purposes only. 

\subsection{Different mass transfer frameworks under isotropic reemission}\label{subsec:models_comparison}

\begin{figure}[!htbp]
    \centering
    \includegraphics[width=\linewidth]{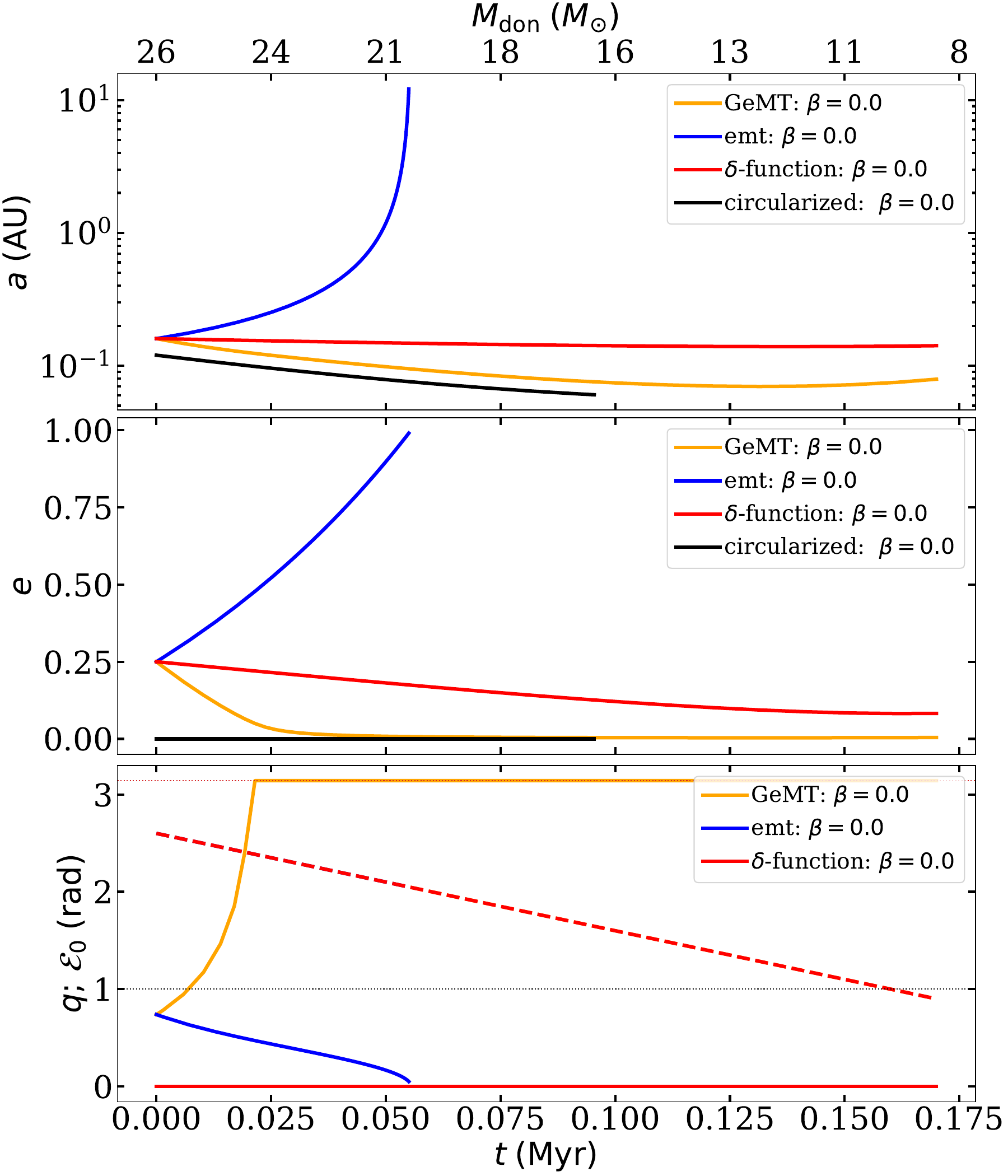}
    \caption{Similar to Fig.~\ref{fig:compare_AML_modes}, but the lines now correspond to different MT frameworks assuming isotropic reemission. The orange, blue, and red lines correspond to the GeMT,  the emt, and the $\delta$-function models, respectively. The black line illustrates the classical RLOF model, meaning instantaneous circularization at periastron $a_{\rm circ} = a(1-e)=0.12$ au and point masses.}
    \label{fig:isot_re_diff_models}
\end{figure}

In Figure~\ref{fig:isot_re_diff_models}, we present the system's evolution predicted by the GeMT model (orange line), alongside results from the $\delta$-function model (red line), the emt model (blue line), and the classical RLOF framework (black line). All models (except for the emt model, which is independent of the AML-parameter $\gamma$) assume isotropic reemission as the AML mode. Within the classical RLOF framework, instantaneous circularization is commonly assumed at the onset of MT. The circularized semimajor axis is typically determined by assuming either (1) conservation of orbital AM, yielding $a_{\rm circ} = a(1-e^2)$ or (2) instantaneous circularization at periapsis, yielding $a_{\rm circ} = a(1-e)$. We include the orbital evolution predicted by the classical framework under assumption 2 (labeled ``circularized'' in Fig.~\ref{fig:isot_re_diff_models}). 

The orbital evolution in the emt model (blue line) qualitatively resembles that of the GeMT model under Jeans mode (blue line in Fig.~\ref{fig:compare_AML_modes}): the orbit widens and the eccentricity increases. As the orbit becomes more eccentric, MT becomes increasingly concentrated around periastron, as shown in the bottom panel of Fig.~\ref{fig:isot_re_diff_models}, where $\mathcal{E}_0 \rightarrow 0$ (blue line). The system evolves until $t \approx 0.055$ Myr, at which point the integration terminates because the orbit becomes effectively parabolic (i.e., $e \geq 0.99$). In contrast, in the $\delta$-function model (red line), the orbit shrinks at the onset of RLOF, and the eccentricity decreases. The subsequent evolution of the orbital elements is relatively weak, yielding a post-MT orbit that is slightly more compact and less eccentric. 

Instantaneous circularization (black line) yields circular and more compact orbits at the onset of RLOF. As expected, when the donor is more massive than the accretor, the orbit shrinks, with the system merging at $t \approx 0.1$ (i.e., $R_{\rm don} \geq  a (1-e$)). In contrast, the GeMT-model evolution (orange line) does result in a merger, illustrating that instantaneous circularization can significantly affect the predicted orbital evolution. Moreover, in the GeMT framework, the parameter $x$ determines whether RLOF occurs for a given mass ratio $q$ and eccentricity $e$ by evolving the system on the $e-x$ plane (e.g., see Fig.~\ref{fig:eccentricity_x_GeMT_AML_modes}). In contrast, the classical RLOF framework does not explicitly consider the radius of the donor during MT. For instance, under circularization 1, the new semimajor axis is $a = 0.15$ au, which corresponds to $x \approx 1.15$, for the GeMT model, this value implies that the donor is detached in the new orbit (i.e., it is located in the no RLOF region in Fig.~\ref{fig:eccentricity_x_GeMT_AML_modes}), and RLOF would not occur. In summary, the GeMT framework does not adopt instantaneous circularization at the onset of RLOF and instead calculates orbital evolution during MT self-consistently.

\section{Summary and discussion}\label{sec:six}

The predicted orbital evolution and final fate of binary systems vary substantially across different MT frameworks. In Sects.~\ref{sec:four} and \ref{subsec:models_comparison}, we compare the predictions of the GeMT framework with those from the $\delta$-function formalism \citep{2007ApJ...667.1170S, 2009ApJ...702.1387S}, the emt model \citep{2019ApJ...872..119H, 2021MNRAS.502.4479H}, and the classical RLOF framework, all under fully nonconservative MT ($\beta = 0$) assuming isotropic reemission. To facilitate comparison, we refer to the already introduced key transitional mass ratios: $q_{\rm trans,a}$, which separates orbital shrinkage from widening, and $q_{\rm trans,e}$, which separates eccentricity damping from pumping. A summary of the key differences is outlined below.

\begin{enumerate}
    \item  For the GeMT model under the point-mass approximation, both transitions occur at $q_{\rm trans,a} = q_{\rm trans,e} = 1.28$, independent of eccentricity $e$. When accounting for extended bodies, the parameter space for orbital widening and eccentricity pumping broadens, and $q_{\rm trans,a} = q_{\rm trans,e}$ becomes eccentricity dependent via the position of the $L_1$ point (global-${L_1}$ fit). Specifically, as eccentricity decreases, both thresholds shift to higher values; conversely, at higher eccentricities, they decrease (Figs.~\ref{fig:colormap_semimajor_axis_ecc} and \ref{fig:colormap_eccentricity_ecc}). 
    \item The $\delta$-function model, also exhibits decreasing $q_{\rm trans,a}$ with increasing $e$, but yields lower thresholds overall. For instance, it predicts orbital widening for $q \leq 1.0$ at $e = 0.99$, and for $q \leq 1.28$ at $e = 0.0$. Eccentricity pumping is predicted for $q \lesssim 1.0$ nearly independent of $e$. Finally, it predicts weaker evolution of both the semimajor axis and eccentricity compared to other frameworks (Figs.~\ref{fig:colormap_semimajor_axis_ecc} and \ref{fig:colormap_eccentricity_ecc}).
    \item The nonconservative extension of the emt model \citep{2021MNRAS.502.4479H} predicts both eccentricity pumping and orbital widening for any mass ratio $q \leq 4$ and eccentricity $e$. Moreover, it predicts stronger evolution in $a$ and $e$ compared to other frameworks (Figs.~\ref{fig:colormap_semimajor_axis_ecc} and \ref{fig:colormap_eccentricity_ecc}).
    \item The classical RLOF framework assumes point masses, synchronous rotation, and instantaneous circularization. Instantaneous circularization yields more compact initial orbits, ensures circular orbits at the end of RLOF, and can lead to mergers in cases where the GeMT model predicts compact and eccentric post-MT orbits (Fig.~\ref{fig:isot_re_diff_models}).
\end{enumerate}

In general, the nonconservative extension of the emt model \citep{2021MNRAS.502.4479H} is independent of the AML-parameter $\gamma$, and generally yields orbital widening and eccentricity growth. Specifically, for $\beta \lesssim 0.1$, the orbit widens and becomes more eccentric for any $0.1 \leq  q \leq 10$ and $ 0.01 \leq e \leq 0.99$, so results obtained with this framework \citep[e.g.,][]{2022ApJ...925..178H,2022MNRAS.517.2111P,2024arXiv241214022P} should be interpreted accordingly. The $\delta$-function model \citep{2007ApJ...667.1170S, 2009ApJ...702.1387S} idealizes MT as an impulsive burst localized at periastron, but hydrodynamical simulations show that even highly eccentric systems do not asymptotically approach the $\delta$-function approximation \citep{2017MNRAS.467.3556B}. Moreover, for systems that initiate MT at low eccentricity, or systems that evolve through a stage with low eccentricity (as we expect them to do, e.g., Fig.~\ref{fig:isot_re_diff_models}), alternative models are required to prevent unphysical negative eccentricities. For example, \cite{2025ApJ...983...39R} transition from the $\delta$-function to the classical RLOF model when $e \leq 0.05$. However, enforcing $e = 0$ eliminates any MT contribution to eccentricity evolution. Lastly and above all, the $\delta$-function model \citep{2007ApJ...667.1170S} as applied in the literature does not conserve orbital AM momentum in the limit of conservative MT (see Sect. 7.3 and Appendix D in \citetalias{2025arXiv250905243P}).\footnote{We resolve this issue by presenting a new set of equations that conserve orbital AM in the limit of conservative MT under a $\delta$-function MT rate centered at periapsis. Nevertheless, the normalized $\delta$-function model can still lead to negative eccentricities, highlighting that this behavior is inherent to the model.} The GeMT framework addresses these issues by providing a unified treatment of conservative and nonconservative MT across arbitrary eccentricities.

\subsection{Eccentric RLOF and formation of GW sources}

Producing binaries that can merge within a Hubble time via GW emission typically requires orbital shrinkage during the MT phase \citep[e.g.,][]{1984ApJ...277..355W,2008ApJS..174..223B,2017MNRAS.471.4256V,2022ApJ...940..184V}. Conventional models often assume instantaneous circularization and consider only circular orbits \citep[e.g.,][]{2021A&A...647A.153B,2021A&A...650A.107M,2021ApJ...922..110G,2024A&A...681A..31P}. In the example presented in Sect.~\ref{subsec:models_comparison} we show that instantaneous circularization and isotropic reemission result in a merger during RLOF (black line in Fig.~\ref{fig:isot_re_diff_models}). In contrast, the GeMT model (orange line in Fig.~\ref{fig:isot_re_diff_models}) does not adopt the instantaneous circularization assumption; the orbital evolution is modeled self-consistently, and it yields a compact post-MT system in a slightly eccentric orbit. In such a system, the donor will collapse, leaving behind a BH. For simplicity, we assume that the orbital parameters do not change during the BH formation. Following \cite{1964PhDT........51P}, we calculate the timescale $t_{\rm merger} = 7.4$ Gyr (or $t_{\rm merger} \approx 0.54$ t$_{\rm Hubble}$) in which the system will merge due to GW radiation. Specifically, 
\begin{equation}\label{eq:gw_timescale}
t_{\mathrm{merge}}(a_0,e_0)
=\frac{12}{19}\,\frac{c_0^4}{\beta}\,\int_{0}^{e_0}
\frac{e^{29/19}\left(1+\frac{121}{304}e^2\right)^{1181/2299}}{(1-e^2)^{3/2}}\,de,
\end{equation}
where
$$
\beta=\frac{64}{5}\,\frac{G^3 m_1 m_2 (m_1+m_2)}{c^5},
\qquad
c_0=\frac{a_0\,(1-e_0^2)}{e_0^{12/19}\left(1+\frac{121}{304}e_0^2\right)^{870/2299}},$$
and $a_0 \approx 0.08$ au, $e_0 \approx 0.005$, $m_{1} = 9$ M$_{\odot}$, and $m_{2} = 10$ M$_{\odot}$. 

\subsection{Occurrence of $L_2$ mass loss}\label{subsec:L_2_occurrence}

Classically, $L_2$ outflow is discussed in the context of contact binaries, \citep[e.g.,][]{1979ApJ...229..223S,2011A&A...528A.114T,2016A&A...588A..50M,2014ApJ...788...22P,2016MNRAS.455.4351P,2017ApJ...850...59P,2024A&A...682A.169H} where mass can escape once the critical equipotential surface through the $L_2$ point is filled. In the context of MT, $L_2$ mass loss is often associated with the onset of unstable MT (but see also \citealt{2024A&A...681A..31P}) as explored in \cite{2015MNRAS.449.4415P,2020ApJS..249....9G,2021A&A...650A.107M},\footnote{Classically, the outer $L_2$ Lagrangian point is defined as the saddle point behind the least-massive object; after mass-ratio reversal, $L_2$ lies behind the donor. Here, we always denote by $L_2$ the Lagrangian point located behind the accretor (Fig.~\ref{fig:L2_mass_loss}).} and in the hydrodynamic simulations of \cite{2018ApJ...868..136M,2018ApJ...863....5M}. This association arises because the mechanism extracts orbital AM very efficiently due to the long lever arm of the $L_2$ point, leading to rapid orbital shrinkage (Figs.~\ref{fig:different_gammas} and \ref{fig:angular_momentum}). Nevertheless, there is observational evidence that some systems can remain stable even after the onset of $L_2$ mass loss; for example, the case of SS433 in our Galaxy \citep[see][]{2004ASPRv..12....1F,2010MNRAS.408....2P}. 

More recently, \cite{2023MNRAS.519.1409L} investigated $L_2$ mass loss in the context of stable MT. They showed that, for mass-transfer rates exceeding $\dot{M}_{\rm don} \gtrsim 10^{-4}$ M$_{\odot} \; \rm yr^{-1}$, it can be energetically favorable for mass to escape via the $L_2$ point rather than through a fast, super-Eddington wind. In this work, we show that for donors synchronous with the orbital velocity at periapsis (i.e., $f_{\rm don} =1 $) and $e \gtrsim 0.2$, the gravitational potential at $L_2$ becomes lower than at $L_1$ across the full range of mass ratios $q$. For supersynchronous donors, this condition holds for all eccentricities, whereas for subsynchronous donors, it occurs only at high eccentricities (see Fig.~\ref{fig:colormap_potential_ratios}). As a result, matter flowing toward the accretor may escape through the $L_2$ point rather than being captured, making MT in eccentric binaries inherently nonconservative due to the evolving potential geometry. Supporting this, \cite{2011ApJ...726...67L} performed hydrodynamical simulations of MT in orbits with $e =0.25$ and found $L_2$ mass loss for both mass ratios $q \sim 1$ and $q \sim 1.7$. 
 
\subsection{Implications of $L_2$ mass loss }\label{subsec:implications_L2}

$L_2$ mass loss may be connected to the observed CB disks around post-red-giant-branch (post-RGB) and post-asymptotic-giant-branch (post-AGB) systems. Interactions between the binary and the disk have been proposed as the driver of the observed eccentricities in these systems. However, \citet{2020A&A...642A.234O} show that binary–disk interactions alone cannot account for the observed eccentric orbits in post-RGB and post-AGB binaries. In fact, such interactions preferentially generate high eccentricities at short orbital periods \citep{2013A&A...551A..50D,2015A&A...579A..49V,2018MNRAS.474..433D}, in contradiction with observations \citep{2024MNRAS.529.3729S}. In \citetalias{2025arXiv250905243P} we show that eccentric RLOF itself serves as an eccentricity pumping mechanism, yielding higher eccentricities at longer orbital periods in agreement with observations. Therefore, since increasing eccentricity makes $L_2$ mass loss more likely, eccentric RLOF potentially leads to the formation of the CB disk in the first place.

The correlation between eccentricity and $L_2$ mass loss has important implications as well for the formation of GW sources. Standard models adopt isotropic reemission as the fiducial AML mode \citep[e.g.,][]{2021ApJ...922..110G,2022ApJ...940..184V,2025ApJ...983...39R}. However, $L_2$ mass loss is expected to contribute in the evolution of the orbital AM (Sect.~\ref{subsec:L_2_occurrence}) and is a far more efficient AML mechanism (see Figs.~\ref{fig:different_gammas}, \ref{fig:compare_AML_modes} and \ref{fig:angular_momentum}). For instance, Fig.~\ref{fig:compare_AML_modes} demonstrates that even $< 10\%$ mass loss via $L_2$ (gray line) results in a tighter post-MT binary than the case of 100\% mass loss through isotropic reemission (orange line). Adopting the resulting orbital parameters and using Eq.~\eqref{eq:gw_timescale} we find $t_{\rm merger} = 0.8$ Gyr or $t_{\rm merger} \approx 0.04$ t$_{\rm Hubble}$. We note that in real systems, it is not necessary that a single AML mechanism operates; instead, a combination of channels might contribute, resulting in a more complicated physical picture. Nevertheless, many binary systems hosting COs are observed with significant eccentricities \citep[e.g.,][]{2024MNRAS.529.3729S,2024OJAp....7E..58E} and even interacting while being eccentric \citep[e.g.,][]{1999AJ....117..587P,2005A&AT...24..151R}, making $L_2$ mass loss an important, and often overlooked, contributor to orbital AM evolution. 

\section{Conclusions}\label{sec:seven}

In this paper, we systematically explored mass transfer in eccentric binaries. We extended the GeMT model introduced in \citetalias{2025arXiv250905243P} to explore the effects of stable, nonconservative mass transfer and its impact on orbital evolution. Since our model builds upon the foundation established in \citetalias{2025arXiv250905243P}, the same limitations and conclusions also apply here. In addition, we presented the first systematic comparison between eccentric mass transfer frameworks; GeMT is the only framework to date that models self-consistent orbital evolution in circular and eccentric orbits under conservative and nonconservative mass transfer, overcoming shortcomings of prior frameworks (Sect.~\ref{sec:six}) Moreover, we demonstrated that the emt, the $\delta$-function (see Appendix D in \citetalias{2025arXiv250905243P}), and the canonical RLOF frameworks are essentially subsets of the GeMT framework.

Under the approximation of a quasi-static gravitational potential \citep{2007ApJ...660.1624S}, we introduced the global-$L_2$ fit, the first prescription for the position of the Lagrangian $L_2$ point (defined here as the Lagrangian point located behind the accretor irrespective of stellar masses) at periapsis, in units of the instantaneous distance between the two stars for asynchronous and eccentric semidetached binaries (Appendix \ref{app:L_2_position}). This fit was used to quantify the angular momentum loss parameter $\gamma$ (i.e., the efficiency of angular momentum extraction during nonconservative mass transfer) for $L_2$ mass loss in circular and eccentric orbits. The specific angular momentum carried away during $L_2$ mass loss is highest for circular orbits and decreases with increasing eccentricity (Fig.~\ref{fig:different_gammas}). Furthermore, increasingly supersynchronous donors (i.e., higher $f_{\rm don}$) result in lower specific angular momentum carried away by the escaping mass (i.e., lower $\gamma$); a trend especially pronounced for systems with less massive donors ($q < 1$, Fig.~\ref{fig:different_gammas}).

We explored orbital evolution under four angular momentum loss modes: (1) Jeans, (2) isotropic reemission, (3) orbital angular momentum loss, and (4) $L_2$ mass loss for circular and eccentric orbits. The orbital evolution of binaries in eccentric orbits differs fundamentally from that in circular orbits. Phase-dependent RLOF drives eccentricity evolution, with higher $e$ producing stronger $\langle \dot{a} \rangle$ at fixed $q$ (see Figs.~\ref{fig:circ_ang_momentum_loss} and \ref{fig:semi_major_beta_0.0}). For a given angular momentum loss mode and accretion efficiency $\beta$, the semimajor axis and eccentricity evolve in a correlated way: orbits either widen while becoming more eccentric, or shrink while circularizing (Figs.~\ref{fig:semi_major_beta_0.0} and \ref{fig:ecc_beta_0.0}). Despite this correlation, different angular momentum loss modes predict substantially different orbital evolutions. In the limit of point masses and fully nonconservative mass transfer ($\beta = 0$), the GeMT model predicts the following according to each mode: 
\begin{enumerate}
    \item Jeans mode. Orbits generally widen and eccentricity increases for nearly all $q$ and $e$. 
    \item Isotropic reemission. Orbits shrink and circularize when $q>1.28$, but widen and become more eccentric when $q<1.28$, independent of $e$. 
    \item Orbital-AML mode. Orbits evolve qualitatively similar to isotropic reemission, but with higher critical mass ratios ($q_{\rm trans,a}=q_{\rm trans,e}=2.0$) and weaker evolution of $a$ and $e$ for $q>2.0$. 
    \item $L_2$ mass loss. Orbits shrink and circularize for $q\gtrsim 0.3$, but widen and become more eccentric when $q \lesssim 0.3$ and $e \gtrsim 0.2$. Overall, $L_2$ mass loss is by far the most efficient mode at extracting orbital angular momentum across most mass ratios for both circular and eccentric orbits. 
\end{enumerate}
In practice, multiple angular momentum loss channels may operate simultaneously; exploring such mixed modes is left for future work.

We demonstrated that eccentric RLOF is an important process for the formation of gravitational-wave sources. By dropping the assumption of instantaneous circularization at the onset of RLOF and modeling orbital evolution self-consistently, we showed that systems that merge during mass transfer under the classical RLOF framework can instead result in compact orbits that produce gravitational-wave sources within a Hubble time (Sect.~\ref{sec:six}). Moreover, we showed that in eccentric orbits, the potential at $L_2$ is lower than at $L_1$ across a wide region of the parameter space (Fig.~\ref{fig:colormap_potential_ratios}); thus, matter flowing toward the accretor may be energetic enough to escape through $L_2$. By quantifying the impact of $L_2$ mass loss in eccentric orbits, we showed that this angular momentum loss channel can contribute significantly to orbital shrinkage, with implications for the formation of gravitational-wave sources. Finally, the correlation between eccentricity and systemic mass loss through the $L_2$ point indicates a possible relation to circumbinary disks observed around eccentric post-RGB and post-AGB binaries (Sect.~\ref{subsec:implications_L2}).  

In summary, the GeMT framework resolves shortcomings of prior frameworks (Sect.~\ref{sec:six}), and unifies them under one general MT framework. The predicted orbital evolution can deviate substantially from classical expectations when dropping the assumption of instantaneous circularization. Incorporating the GeMT model into binary evolution and population synthesis codes would enable more accurate and self-consistent tracking of the orbital evolution for both circular and eccentric systems under conservative and nonconservative mass transfer, thereby refining our understanding of stable mass transfer in shaping the evolution of interacting binaries with important implications from main-sequence mass transfer to the formation of gravitational-wave sources.

\section{Data availability}

The data necessary to reproduce Figs. 10-13 in this paper is available on \href{https://doi.org/10.5281/zenodo.18151287}{Zenodo}. The GeMT code will be shared upon reasonable request to the authors.

\begin{acknowledgements}
The authors would like to thank the referee for their constructive feedback. AP would like to thank Floris Kummer, Caspar Bruenech, and Thomas Tauris for the useful discussions. AP \& ST acknowledges support from the Netherlands Research Council NWO (VIDI 203.061 grant). EL acknowledges support through a start-up grant from the Internal Funds KU Leuven (STG/24/073) and through a Veni grant (VI.Veni.232.205) from the Netherlands Organization for Scientific Research (NWO). FD acknowledges support from the UK’s Science and Technology Facilities Council Grant No. ST/V005618/1. This work used the following software packages: Matplotlib \cite{Hunter:2007}, NumPy \cite{harris2020array}, SciPy \citep{2020SciPy-NMeth}, SymPy \citep{10.7717/peerj-cs.103} and PySR \citep{2023arXiv230501582C}.
\end{acknowledgements}

\bibliographystyle{aa}
\bibliography{references}

\begin{appendix}

\section{Potential height at the $L_2$ Lagrangian point in eccentric orbits}\label{app:potential_height}

Traditionally, the locations of the Lagrangian points $L_1 - L_5$ are defined under the assumption of circular orbits with synchronously rotating component stars. However, in eccentric binaries, the orbital angular velocity varies with time, making synchronous rotation throughout the orbit impossible. Eccentricity and/or asynchronous donor rotation, therefore, yields a time-varying gravitational potential $V$.  \citet{2007ApJ...660.1624S} showed, however, that the potential can be treated as quasi-static, provided the donor’s dynamical timescale is much shorter than the orbital and rotational timescales--a condition known as the first approximation \citep{1963ApJ...138.1112L}. Under this approximation, the donor's structure is assumed to adjust instantaneously to the changing potential throughout the orbit. 

Following \cite{2007ApJ...660.1624S}, the quasi-static potential is written as 
\begin{equation}\label{eq:potential}
    V = X - \frac{q}{X} - \frac{1}{\sqrt{r^2-2X+1}} - \frac{1}{2}(X^2 + Y^2)(1+q)\mathcal{A}(f_{\rm don},e,\mathcal{E}),
\end{equation}
where $X$ and $Y$ are the Cartesian coordinates in the plane of the orbit, expressed in units of the instantaneous orbital separation (i.e., $X \equiv X/r$ and $Y \equiv Y/r$),
and 
\begin{equation}\label{eq:A_function}
    \mathcal{A}(f_{\rm don},e,\mathcal{E})= f_{\rm don}^2 \frac{1+e}{(1-e)^3} \biggl(\frac{r}{a} \biggr)^3,
\end{equation}
quantifies the deviation of the donor's spin velocity from the orbital angular velocity at periapsis as a function of the eccentricity $e$ and eccentric anomaly $\mathcal{E}$\footnote
{We note that we choose the eccentric anomaly to express $\mathcal{A}(f_{\rm don},e,\mathcal{E})$ for consistency with \citetalias{2025arXiv250905243P}, however other angle parameters, such as true anomaly, are equally valid.}. We determine quasi-static Lagrangian points $L_1 -L_5$ by solving $\nabla V =0$ with respect to the coordinates $X$ and $Y$. Throughout this work, we define the Lagrangian point located behind the accretor as $L_2$ and the one behind the donor as $L_3$, irrespective of the stellar masses.

\begin{figure}[!htbp]
\includegraphics[width=\linewidth]{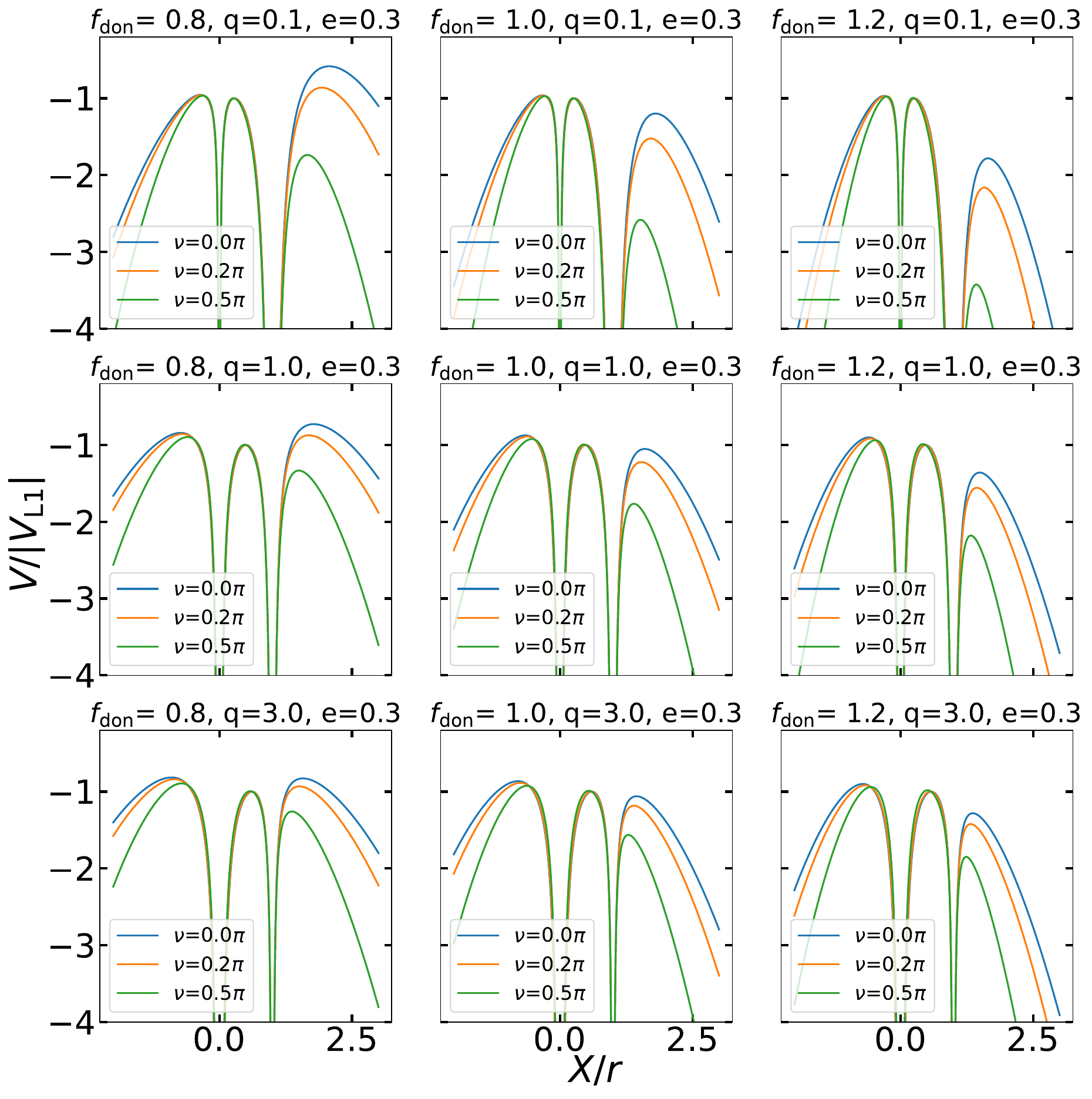}
\caption{Height of the potential $V$ along the axis connecting centers of mass of the binary components (X-axis). The potential is normalized to its value at the inner Lagrangian point $L_1$. For all models $e=0.3$. The blue, orange, and green lines correspond to orbital phases $\nu = 0.0 \pi, \pi/4,$ and $ \pi/2$, respectively. From top to bottom, the subfigures correspond to $q=0.1, 1.0,$ and $3.0$, respectively. From left to right, the subfigures correspond to $f_{\rm don} =0.8, 1.0,$ and $1.2$, respectively.}
\label{fig:potential_heights_true_anomaly}
\end{figure}

In Figure~\ref{fig:potential_heights_true_anomaly}, we show the potential $V$ at the collinear Lagrangian points as a function of orbital phase, normalized to its value at the inner Lagrangian point $L_1$. Among these points, the potential at $L_2$ exhibits the strongest dependence on orbital phase, peaking at periapsis ($\nu = 0 \pi$; blue line) and decreasing throughout the rest of the orbit. Moreover, the position of the $L_2$ point reaches its maximum distance from the donor at periapsis, while it moves closer at later orbital phases. These trends are qualitatively consistent across the entire parameter space. Additionally, the relative potential height at $L_2$ decreases with increasing $f_{\rm don}$ (i.e., more supersynchronous donors) and increasing mass ratio $q$ (i.e., more massive donors), while at $L_3$ it remains largely constant over the considered range of parameters. Finally, the potential height at $L_2$ can be lower than at $L_1$ across a wide region of the parameter space, while at $L_3$ is always higher than at $L_1$.

\begin{figure}[!htbp]
    \centering
    \includegraphics[width=\linewidth]{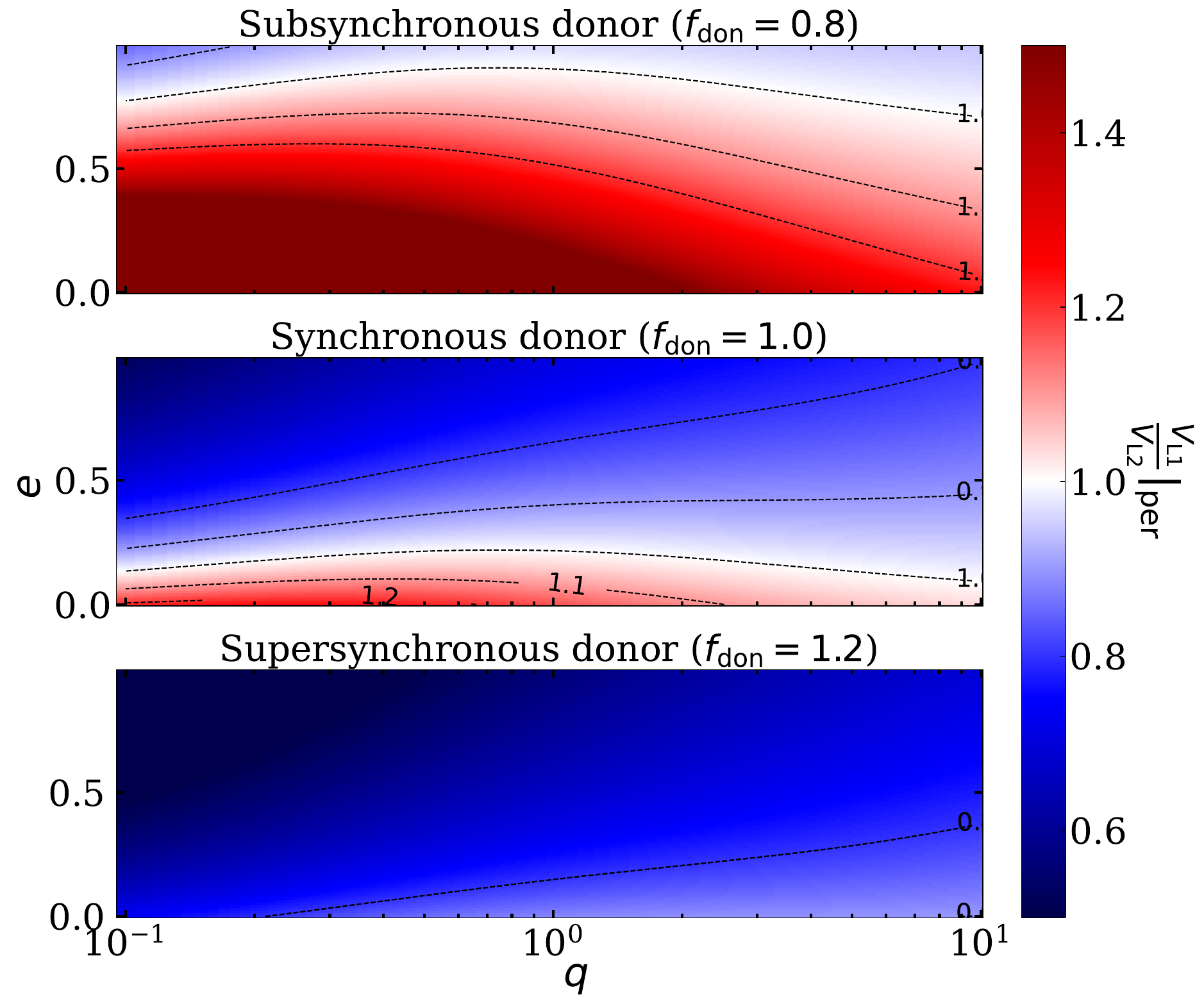}
    \caption{Height of the potential at $L_1$ divided by that at $L_2$ at the periapsis of the orbit as a function of mass ratio, $q$, and eccentricity, $e$. From top to bottom, the donor's rotational factor is $f_{\rm don}= 0.8, 1.0$ and $1.2$, respectively.}
    \label{fig:colormap_potential_ratios}
\end{figure}

In Figure~\ref{fig:colormap_potential_ratios}, we show the ratio of the potential height at $L_1$ to that at $L_2$ at the periapsis of the orbit for $f_{\rm don}= 0.8, 1.0$ and $1.2$, respectively.  Blue regions (i.e., $V_{\rm L1}/V_{\rm L2} < 1.0$) indicate that the potential at $L_2$ is lower than at $L_1$, while red regions (i.e., $V_{\rm L1}/V_{\rm L2} > 1.0$) indicate the opposite. The color intensity reflects the magnitude of the ratio.

The potential height at $L_2$ relative to that at $L_1$ decreases (i.e., $V_{\rm L1}/V_{\rm L2}$ decreases) for increasing $e$. Notably, for any $f_{\rm don} \geq 1.0$ (i.e., synchronous or supersynchronous donors with the orbital velocity at periapsis) and $e \geq 0.2$, the potential height at $L_2$ is lower than at $L_1$ across the full range of mass ratios $q$. Consequently, for binaries that undergo RLOF with such parameter--combinations corresponding to the blue regions in Fig.~\ref{fig:colormap_potential_ratios}--matter flowing towards the accretor might not necessarily be captured and may be energetic enough to escape the system via the $L_2$ point. Importantly, even in systems corresponding to the red regions, $L_2$ mass loss may still occur: although the potential at $L_2$ is initially higher than at $L_1$ near periapsis, this relationship can reverse later at later orbital phases, with the potential at $L_2$ becoming lower than at $L_1$ (see Fig.~\ref{fig:potential_heights_true_anomaly}). These results suggest that MT in eccentric binaries may be inherently nonconservative due to the evolving geometry of the potential.

\section{Angular momentum loss via $L_2$ mass loss}\label{app:L_2_position}

In the center-of-mass frame, see Fig.~\ref{fig:L2_mass_loss}, the position of the $L_2$ point determines the length of the lever arm and thus the specific angular momentum that is lost in the case of $L_2$ mass loss. The position, $X$, of the $L_2$ point--relative to the donor's center of mass--can be found by solving
\begin{equation}\label{eq:Lagrangian_points}
    \frac{q}{X_{\rm L}^2} + \frac{1}{(1-X_{\rm L}^2)} - X_{\rm L}(1+q)\mathcal{A}(f_{\rm don},e,\mathcal{E}) +1 = 0,
\end{equation}
where $X_{\rm L}$ is expressed in units of the instantaneous orbital separation (i.e., $X_{\rm L} = X/r$). Following the same approach as in \citetalias{2025arXiv250905243P}, we solve for the periapsis of the orbit  by evaluating $\mathcal{A}(f_{\rm don},e,\mathcal{E} =0)$, and thus Eq.~\eqref{eq:Lagrangian_points} is written as
\begin{equation}\label{eq:Lagrangian_points_2}
    \frac{q}{X_{\rm L}^2} + \frac{1}{(1-X_{\rm L}^2)} - X_{\rm L}(1+q)f_{\rm don}^2(1+e) +1 = 0.
\end{equation}
We fit the prescription
\begin{flalign}\label{eq:fit_formula_L2}
    X_{\rm L2}&(f_{\rm don},q,e) = 1.057f_{\rm don}^{-0.035} \nonumber \\ 
    &+\biggr(f_{\rm don}(1.895e+1.437f_{\rm don})(f_{\rm don} (1+q)-0.233)\biggl)^{-0.459}
\end{flalign}
to the numerical solutions of Eq.~\eqref{eq:Lagrangian_points_2}, such that $X_L = X_{\rm L2}(f_{\rm don},q,e)$. The fit determines the position of the $L_2$ point at periapsis in units of the instantaneous distance between the two stars, and achieves an accuracy better than $\sim 7\%$ for $0.5 \leq f_{\rm don} \leq 2.0$, $0.1 \leq q \leq 10.1$ and $0.0 \leq e \leq 0.99$.

\begin{figure}[!htbp]
    \centering
    \includegraphics[width=\linewidth]{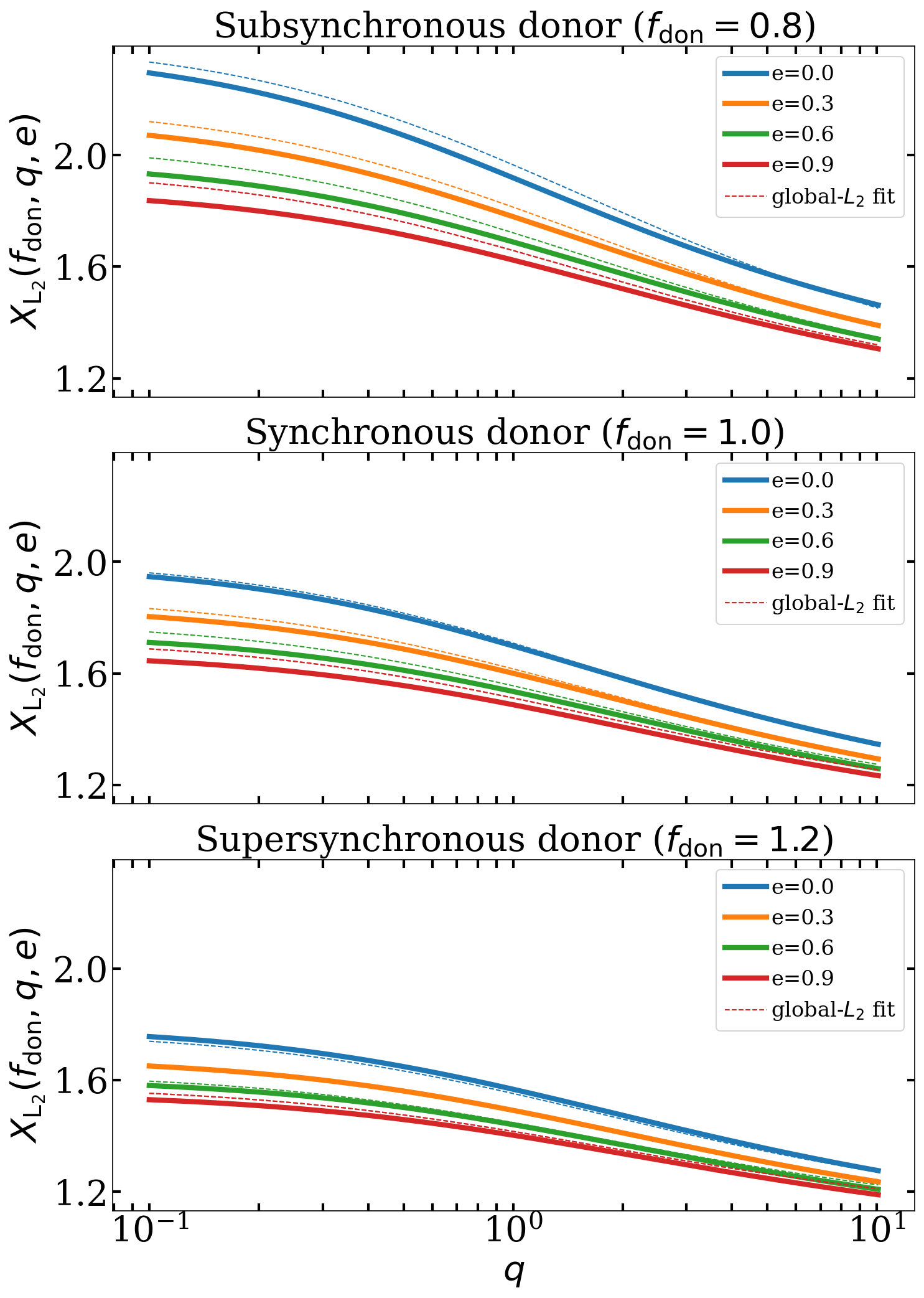}
    \caption{Position of the $L_2$ point, relative to the donor's center of mass, at the periapsis of the binary orbit in units of the instantaneous binary separation. From top to bottom, the subfigures correspond to $f_{\rm don}=0.8,1.0,$ and $1.2$, respectively. The thick lines are the numerical solutions of Eq.~\eqref{eq:Lagrangian_points_2}. Blue, orange, green, and red colors correspond to $e=0.0,0.3,0.6,$ and $0.9$, respectively. The dashed lines correspond to Eq.~\eqref{eq:fit_formula_L2}.}
    \label{fig:fit_L2}
\end{figure}

In Figure~\ref{fig:fit_L2}, we compare the global-$L_2$ fit (i.e., Eq.~\ref{eq:fit_formula_L2}) predictions to the numerical solutions of Eq.~\eqref{eq:Lagrangian_points_2} for varying eccentricity $e$ and donor spins $f_{\rm don}$, across different mass ratios $q$. An increase in both $q$ or $e$ brings the position of the $L_2$ point closer to the donor's center of mass. Conversely, lower donor spins move the $L_2$ point further from the donor. Finally, we emphasize that although $X_{\rm L}$ is not a function of the orbital phase, the location of $L_2$ itself, $X = X_{\rm L2}(f_{\rm don},q,e) r$ is phase-dependent.

We utilize the global-$L_2$ fit to determine the length of the lever arm $D_{\rm L2}$ and thus the specific angular momentum that is carried away in the case of $L_2$ mass loss. From Fig.~\ref{fig:L2_mass_loss}, we observe that in the center-of-mass frame
\begin{flalign}\label{eq:deriv_gamma}
    \dot{J}_{\rm orb,ml}/J_{\rm orb} &= \Omega_{\rm orb} D_{\rm L2}^2 \nonumber \\
     & =\Omega_{\rm orb} \biggr(X - D_{\rm don} \biggl)^2  \nonumber \\
     & = \sqrt{\frac{GM}{a^3}} \sqrt{1-e^2}\biggr(\frac{a}{r}\biggl)^2 \biggr((X_{\rm L2}(f_{\rm don},q,e) - \frac{1}{1+q})r \biggl)^2 \nonumber \\
     & = \sqrt{GMa(1-e^2)} (X_{\rm L2}(f_{\rm don},q,e) - \frac{1}{1+q})^2 \nonumber \\
    & = J_{\rm orb} \frac{M}{M_{\rm acc} M_{\rm don}} (X_{\rm L2}(f_{\rm don},q,e) - \frac{1}{1+q})^2 \nonumber \\
    & =  \frac{M^2}{M_{\rm acc} M_{\rm don}} (X_{\rm L2}(f_{\rm don},q,e) - \frac{1}{1+q})^2 \frac{J_{\rm orb}}{M} \nonumber \\
    & =  (2+q+\frac{1}{q}) (X_{\rm L2}(f_{\rm don},q,e) - \frac{1}{1+q})^2 \frac{J_{\rm orb}}{M}, 
\end{flalign}
and, combining Eqs.~\eqref{eq:angular_momentum_parametrization} and \eqref{eq:deriv_gamma}, we arrive at Eq.~\eqref{eq:gamma_L2}. 

\begin{figure}[!htbp]
    \includegraphics[width=\linewidth]{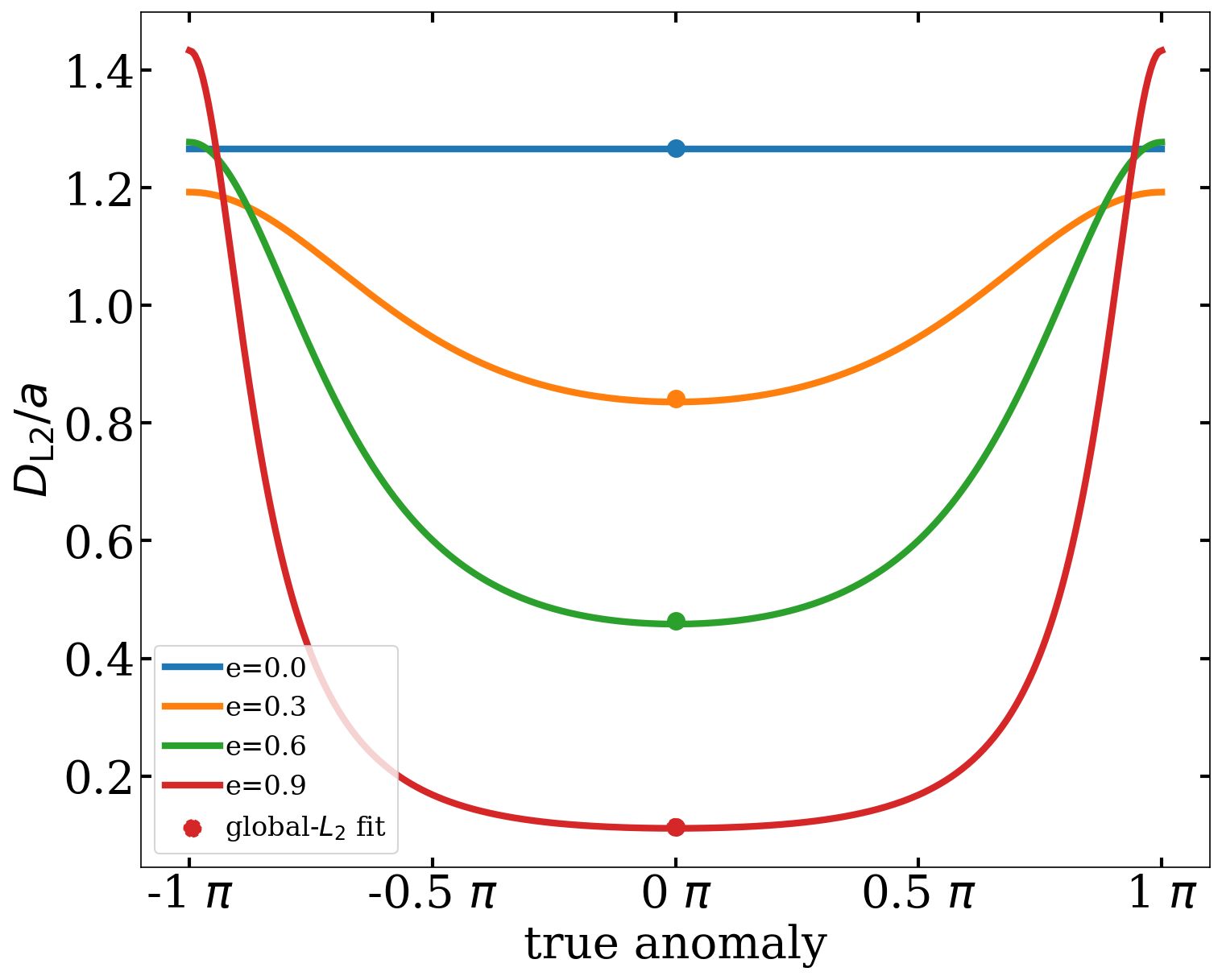}
    \caption{Position of the $L_2$ Lagrangian point, relative to the system's center of mass, in units of the semimajor axis, for $q=3.0$ and $f_{\rm don} =1.0$ as a function of true anomaly. The thick lines are calculated based on the numerical solutions of Eq.~\eqref{eq:Lagrangian_points}. The circular points are calculated using the global-$L_2$ fit (Eq.~\ref{eq:fit_formula_L2}). Blue, orange, green, and red colors correspond to $e=0.0,0.3,0.6,$ and $0.9$, respectively.}
    \label{fig:L2_anal_func_error}
\end{figure}

The key advantage of our approach is that it yields a phase-independent AML-parameter $\gamma$ for $L_2$ mass loss, making it particularly suitable for secular orbital evolution models. In Fig.~\ref{fig:L2_anal_func_error}, we present the length of the lever arm $D_{\rm L2}$, derived from the numerical solutions of Eq.~\eqref{eq:Lagrangian_points}, over one orbit and varying eccentricity $e$, expressed in units of the semimajor axis $a$. We also compare these results to predictions using the global-$L_2$ fit model. Our approximate solution is very accurate at periapsis. Moreover, the $L_2$ point reaches its minimum distance from the system's center of mass at periapsis; thus, Eq.~\eqref{eq:deriv_gamma} yields the minimum orbital AML for a given set of parameters $f_{\rm don}$, $q$, and $e$. 

In summary, we use the global-$L_2$ fit to evaluate the AML during $L_2$ mass loss (Eq.~\ref{eq:deriv_gamma}). Moreover, the global-$L_2$ fit is evaluated at periapsis mainly for two reasons: (1) for nonzero eccentricities, the MT rate in the GeMT model peaks at periapsis and declines at later orbital phases (see Sect. 3.4 in \citetalias{2025arXiv250905243P}), and (2) the GeMT model assumes instantaneous MT from the donor to the accretor, as well as instantaneous mass loss from the system in the case of nonconservative MT. This approach also ensures consistency with the global-$L_1$ fit described in \citetalias{2025arXiv250905243P}. Nevertheless, we emphasize that alternative prescriptions for the $L_2$ position may also be valid, depending on assumptions regarding the orbital phase at which mass loss occurs.

\section{Explicit expressions for the functions appearing in the orbit-averaged equations of motion}\label{app:dimensionless_functions}

In this section, we explicitly present the dimensionless functions $f_{\dot{M}_{\rm don}}(e,x), \; f_{a}(e,x), \; f_{e}(e,x), \; g_{a}(e,x), \; g_{e}(e,x), \; h_{a}(e,x),$ and $ \; h_{e}(e,x)$, as referenced in Sect.~\ref{sec:one}.

\begin{flalign}
f_{\dot{M}_{\rm don}}&(e,x) = - \frac{1}{96 \pi} \Biggl(36 e^4 x^3 \mathcal{E}_0 + 3e^4  \nonumber x^3\sin{(4\mathcal{E}_0)} \\ \nonumber
&- 32e^3 x^3 \sin{(3\mathcal{E}_0)} + 24 e^3 x^2 \sin{(3\mathcal{E}_0)} \\ \nonumber
&+ 288 e^2 x^3 \mathcal{E}_0 - 432 e^2 x^2 \mathcal{E}_0   \\ \nonumber
&+24 e^2 x(e^2 x^2 +6x^2 -9x +3)\sin(2\mathcal{E}_0) +144 e^2 x \mathcal{E}_0  \\ \nonumber
&-24 e(12e^2x^3-9e^2x^2+16x^3-36x^2+24x-4)\sin{(\mathcal{E}_0)} \\ 
&+96x^3 \mathcal{E}_0 -288x^2\mathcal{E}_0 +288x\mathcal{E}_0 -96 \mathcal{E}_0\Biggr), \\ \nonumber
\end{flalign}

\begin{flalign}
f_{a}&(e,x) =  \frac{1}{96 \pi} \Biggl(36 e^4 x^3 \mathcal{E}_0 + 3e^4 x^3 \nonumber \sin{(4\mathcal{E}_0)} \\ \nonumber
&- 16 e^3 x^3 \sin{(3\mathcal{E}_0)}  + 24 e^3 x^2 \sin{(3\mathcal{E}_0)} - 144 e^2 x^2 \mathcal{E}_0 \\ \nonumber
&+ 24 e^2 x(e^2 x^2 -3x +3) \sin(2\mathcal{E}_0) +144 e^2 x \mathcal{E}_0    \\ \nonumber
&+24 e (-6 e^2 x^3 +9 e^2 x^2 + 8 x^3 -12 x^2 +4)\sin{(\mathcal{E}_0)} \\ 
&-96x^3 \mathcal{E}_0 +288x^2\mathcal{E}_0 -288x\mathcal{E}_0 +96 \mathcal{E}_0\Biggr),  \\ \nonumber
\end{flalign}

\begin{flalign}
f_{e}&(e,x) =  - \frac{e^{2} - 1}{32 \pi} \Biggl(12 e^{3} x^{3} \nonumber \mathcal{E}_0 + e^{3} x^{3} \sin{(4 \mathcal{E}_0 )} \\ \nonumber
&- 8 e^{2} x^{3} \sin{(3 \mathcal{E}_0 )} + 8 e^{2} x^{2} \sin{(3 \mathcal{E}_0 )} + 48 e x^{3} \mathcal{E}_0 \\ \nonumber
&- 96 e x^{2} \mathcal{E}_0 + 8 e x (e^{2} x^{2} + 3 x^{2} - 6 x + 3) \sin{(2 \mathcal{E}_0 )} \\ 
&+ 48 e x \mathcal{E}_0  \\ \nonumber
&- (72 e^{2} x^{3} - 72 e^{2} x^{2} + 32 x^{3} - 96 x^{2} + 96 x - 32) \sin{(\mathcal{E}_0 )}\Biggr), \\ \nonumber
\end{flalign}

\begin{flalign}
g_{a}&(e,x) =\frac{1}{32 \pi} \Biggl( ex 
\Bigl( - 16(6+e^{2}(x-3) - 4x)x\sin{(\mathcal{E}_0)} \\
&+ e^{2}x [16(1- x) \sin{(3 \mathcal{E}_0 )} + 3 e x \sin{(4 \nonumber \mathcal{E}_0)}] \\ \nonumber 
&+ 8e(((e^{2} + 2)x - 6)x + 3) \sin{(2 \mathcal{E}_0)}\Bigr)  \\ \nonumber
&+ 4 \mathcal{E}_0 x(-8(3+(x-3)x)+e^{2}(12+(e^{2}-8)x^{2})) \\ 
&+ 64 \sqrt{1 - e^{2}} \operatorname{asin}{(\frac{\sqrt{e + 1} \sqrt{\frac{1}{1 - e}} \sin{(\frac{\mathcal{E}_0}{2} )}}{\sqrt{\cos^{2}{(\frac{\mathcal{E}_0}{2} )} + \frac{(e + 1) \sin^{2}{(\frac{\mathcal{E}_0}{2} )}}{1 - e}}} )}\Biggr), \\  \nonumber
\end{flalign}

\begin{flalign}
g_{e}&(e,x) = \frac{(1 - e^{2})}{48 \pi e} \Biggl(e^{2} x \Bigl(e x [3 e x \sin{(4 \nonumber \mathcal{E}_0 )} \\ \nonumber 
&+ (18 - 20 x) \sin{(3 \mathcal{E}_0 )}] \\ \nonumber 
&+ 6([2 (e^{2} + 4)x - 15]x + 6) \sin{(2 \mathcal{E}_0 )}\Bigr) \\ \nonumber 
&- 6 e \Bigl([e^{2} x (14 x - 15) +  8\bigl((x - 3)x + 3\bigr)]x - 4\Bigr) \sin{(\mathcal{E}_0 )} \\ \nonumber 
&+ 12\Bigl(e^{2} x \bigr([(e^{2} + 4)x - 9]x + 6 \bigl) - 2 \Bigr) \mathcal{E}_0 \\  
&+ 48 \sqrt{1 - e^{2}} \operatorname{asin}{(\frac{\sqrt{e + 1} \sqrt{\frac{1}{1 - e}} \sin{(\frac{\mathcal{E}_0}{2} )}}{\sqrt{\cos^{2}{(\frac{\mathcal{E}_0}{2} )} + \frac{(e + 1) \sin^{2}{(\frac{\mathcal{E}_0}{2} )}}{1 - e}}} )}\Biggr), \\ \nonumber
\end{flalign}

\begin{flalign}
h_{a}&(e,x) =\frac{1}{8 \pi} \Biggl( -4 \mathcal{E}_0 x \nonumber
\Bigl(6+x(-6+(2+e^{2})x)\Bigr)  \\ \nonumber
&+ 2e \Bigl((-12x-(-4+e^{2})x^{3}-\frac{4}{-1+e\cos{\mathcal{E}_0}} \Bigr)\sin{(\mathcal{E}_0}) \\  \nonumber 
&+ ex^{2}(3(-2+x)\sin{(2\mathcal{E}_0}) -ex\sin{(3\mathcal{E}_0}))\Bigr) \\  
&+\frac{16}{\sqrt{1-e^2}} \operatorname{asin}{(\frac{\sqrt{e + 1} \sqrt{\frac{1}{1 - e}} \sin{(\frac{\mathcal{E}_0}{2} )}}{\sqrt{\cos^{2}{(\frac{\mathcal{E}_0}{2} )} + \frac{(e + 1) \sin^{2}{(\frac{\mathcal{E}_0}{2} )}}{1 - e}}} )}\Biggr), \\  \nonumber  
\end{flalign}

and

\begin{flalign}
h_{e}&(e,x) =\frac{1}{48 e\pi} \Biggl( \frac{1} {-1+e\cos{(\mathcal{E}_0})}(-1+e^{2})  \nonumber \\  
&\Bigl[12 \mathcal{E}_0 (-2+e^{2}x^{2}(-3+2x)) \nonumber \\ 
&+ 4e \Bigl(3\mathcal{E}_0(2+e^{2}(3-2x)x^{2})\cos{(\mathcal{E}_0}) +6\sin{(\mathcal{E}_0}) \nonumber \\
&+x(-1+e\cos{(\mathcal{E}_0}))\Bigl(4(9+x(-9+(3+2e^{2})x))  \nonumber\\
&+ ex(9(3-2x)\cos{(\mathcal{E}_0)}+4ex\cos{(2\mathcal{E}_0}))\Bigr)\sin{(\mathcal{E}_0})\Bigl)\Bigl] \nonumber \\
&+48\sqrt{1-e^2}\operatorname{asin}{(\frac{\sqrt{e + 1} \sqrt{\frac{1}{1 - e}} \sin{(\frac{\mathcal{E}_0}{2} )}}{\sqrt{\cos^{2}{(\frac{\mathcal{E}_0}{2} )} + \frac{(e + 1) \sin^{2}{(\frac{\mathcal{E}_0}{2} )}}{1 - e}}} )}\Biggr).  \\   \nonumber 
\end{flalign}
\end{appendix}

\end{document}